\documentclass[]{interact}

\usepackage{epstopdf}
\usepackage[caption=false]{subfig}
\usepackage{float}
\usepackage[numbers,sort&compress]{natbib}
\usepackage[colorlinks=true, citecolor=blue, urlcolor=blue]{hyperref}
\bibpunct[, ]{[}{]}{,}{n}{,}{,}
\makeatletter
\def\NAT@def@citea{\def\@citea{\NAT@separator}}
\makeatother

\theoremstyle{plain}

\theoremstyle{definition}

\theoremstyle{remark}

\begin{document}

\title{Quantum Control in Open and Periodically Driven Systems}
\author{
\name{Si-Yuan Bai \textsuperscript{a}, Chong Chen \textsuperscript{b}, Hong Wu \textsuperscript{a}, and Jun-Hong An \textsuperscript{a, $\dag$} \footnote{$\dag$ Email: anjhong@lzu.edu.cn}}
\affil{\textsuperscript{a} School of Physical Science and Technology \& Key Laboratory for Magnetism and Magnetic Materials of the MoE, Lanzhou University, Lanzhou 730000, China\\
\textsuperscript{b} Department of Physics and The Hong Kong Institute of Quantum Information of Science and Technology, The Chinese University of Hong Kong, Shatin, New Territories, Hong Kong, China}
}

\maketitle

\begin{abstract}
Quantum technology resorts to efficient utilization of quantum resources to realize technique innovation. The systems are controlled such that their states follow the desired manners to realize different quantum protocols. However, the decoherence caused by the system-environment interactions causes the states deviating from the desired manners. How to protect quantum resources under the coexistence of active control and passive decoherence is of significance. Recent studies have revealed that the decoherence is determined by the feature of the system-environment energy spectrum: Accompanying the formation of bound states in the energy spectrum, the decoherence can be suppressed. It supplies a guideline to control decoherence. Such idea can be generalized to systems under periodic driving. By virtue of manipulating Floquet bound states in the quasienergy spectrum, coherent control via periodic driving dubbed as Floquet engineering has become a versatile tool not only in controlling decoherence, but also in artificially synthesizing exotic topological phases. We will review the progress on quantum control in open and periodically driven systems. Special attention will be paid to the distinguished role played by the bound states and their controllability via periodic driving in suppressing decoherence and generating novel topological phases.
\end{abstract}

\begin{keywords}
Decoherence control; Bound state; Topological phases; Floquet engineering
\end{keywords}
\begin{figure}[H]
\includegraphics[height=0.7\textwidth]{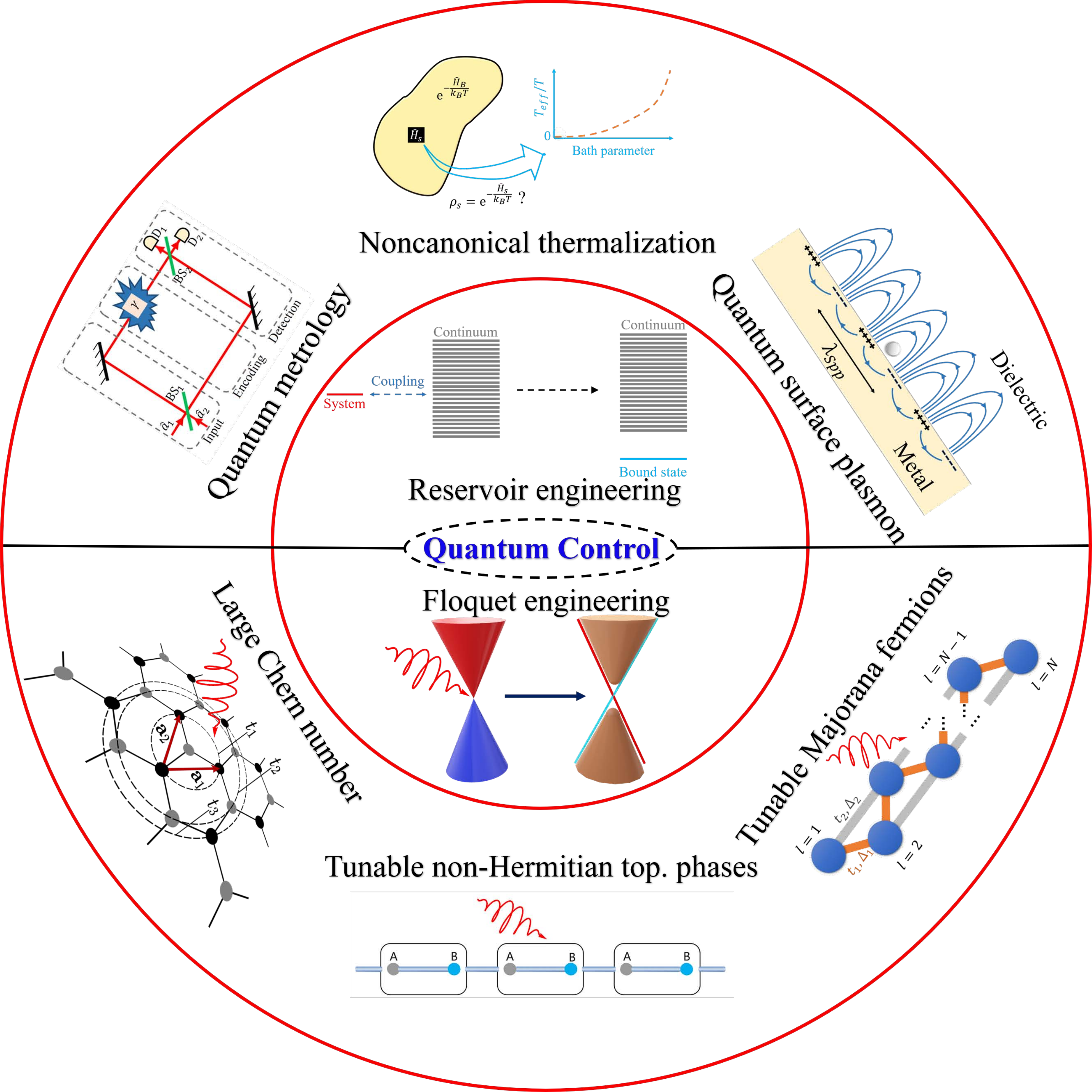}
\end{figure}

\section{Introduction}
Quantum technology is a class of technology that works by using quantum effects, such as quantum coherence, squeezing, and entanglement, to develop revolutionary techniques. Including quantum computation and algorithm \cite{Feynman1982,Ekert1996,Law2001,DiVincenzo2000}, quantum metrology and sensor \cite{Giovannetti2004,Leibfried2004,Giovannetti2011,Escher2011}, quantum simulation \cite{Georgescu2014}, and quantum communication \cite{Bennett1992,Mattle1996}, it opens an avenue to break through the technique limits for their classical counterparts from the principles of quantum mechanics. The main idea is to actively control the quantum states of relevant systems in the desired manners to realize different protocols. However, its realization is challenged by decoherence \cite{Zurek2003}. As a ubiquitous phenomenon in the microscopic world, decoherence is caused by the inevitable interaction of quantum systems with their environments. It makes the quantum resources degraded and the states deviating from the desired manners. Therefore, a thorough understanding of the detrimental influences of decoherence on quantum protocols and their efficient control are of importance in quantum technology.
Many methods, such as entanglement concentration \cite{Bennett1996,Zhao2001}, dynamical decoupling \cite{Viola1998,Lange2010}, and feedback control \cite{Lloyd2000a,Tombesi1995}, have been proposed to suppress the effects of decoherence. Recent studies show that the structured environment, e.g. photonic crystal \cite{John1990, Ogawa2004, Fujita2005, Jorgensen2011} and localized surface plasmons \cite{Chang2006, Tame2013}, dramatically changes or even inhibits the decoherence of quantum emitters \cite{Ogawa2004, Yang2017, Yang2019}. Further studies reveal that these phenomena are due to the formation of a bound state in the energy spectrum of the total system \cite{Miyamoto2005}. It has been reported that such a bound state has profound impacts on many quantum protocols and nonequilibrium physics, e.g. entanglement preservation \cite{Tong2010,Lue2013,Yang2013}, noncanonical thermalization \cite{Yang2014}, quantum speed limit \cite{Liu2016}, and quantum metrology \cite{Wang2017,Bai2019,Wu2020a}. The results give us an insightful guideline to control decoherence via forming the bound state by quantum reservoir engineering \cite{Poyatos1996,Metelmann2015}.

On the other hand, coherent control via periodic driving dubbed as Floquet engineering has become a versatile tool in quantum control \cite{Oka2019}. The main idea resides in that the concepts of the energy spectrum and bound states of static systems can be completely inherited in the quasienergy spectrum of periodically driven systems according to Floquet theory. The bound-state mechanism in static systems was generalized to the periodically driven systems as the Floquet bound-state mechanism \cite{Chen2015,Ma2018,Yang2019a,Bai2020}. Periodic driving largely extends the controllability of systems because time as a new control dimension is introduced. It overcomes the practical difficulty of reservoir engineering in static systems that the parameter is hard to change once the material is fabricated \cite{Grifoni1998,Thorwart2000,Kohler2005}. Engineering the quasienergy spectrum to form the Floquet bound states via periodic driving can also be applied to artificially synthesizing topological phases. The periodic driving induces an effective long-range hopping in lattice systems which is hard to experimentally realize in the static system \cite{Lindner2011,Tong2013,Xiong2016,Liu2019}. In this manner, widely tunable numbers of edge or surface states have been realized by periodic driving \cite{Rudner2013,Kundu2014,Tong2013,Xiong2016,Liu2019,Lababidi2014,Wu2020}. It is found that not only the topological phases hard to reach in the same setting of static systems but also many exotic topological phases totally absent in static systems can be realized by Floquet engineering \cite{Cayssol2013,Rudner2013,Mukherjee2017}.

We here review the progress on quantum control in open and periodically driven systems. The paper is organized as follows. In Sec. \ref{sec:decoh}, the bound-state mechanism is given and the decoherence control based on the reservoir engineering is reviewed. The profound impacts of the bound state on entanglement preservation, quantum speedup, noncanonical thermalization, quantum metrology, and quantum plasmonics are explicitly analyzed. The Floquet engineering is reviewed in Sec. \ref{sec:Floquet}. The applications of Floquet engineering in decoherence control and the generation of exotic topological phases are explicitly discussed. Conclusion and outlook are made in Sec. \ref{sec:conclu}.

\section{Decoherence control in open quantum systems}\label{sec:decoh}
Quantum technology is based on the efficient utilization of quantum resources, e.g. quantum coherence and entanglement, of the relevant systems to realize the technique innovation. However, any microscopic system is inevitably influenced by its surrounding environment and becomes an open system. It causes unwanted loss of quantum resources. Therefore, one of the important purposes of quantum control in open systems is to suppress decoherence. Depending on whether the open system has energy exchange with the environment, its decoherence can be classified into dissipation and dephasing. We focus on dissipation.

Open system influenced by its environment has a wide physical relevance in different experimental platforms of quantum technology. There are two types of open systems: Discrete- and continuous-variable systems. The discrete-variable system is generally a two-level system, which may be a two-level atom interacting with the radiation electromagnetic field \cite{Scully1997}, an electron spin in a quantum dot interacting with the nuclear spin \cite{Khaetskii2002,Merkulov2002}, or a superconducting charge qubit exposed to the charge noise \cite{Bouchiat1998}. The continuous-variable one may be a harmonic oscillator or a single-mode quantized optical field.  The microscopic description of dissipation is based on the ``system" + ``reservoir'' approach pioneered by Senitzky \cite{Senitzky1960,Senitzky1961}. The Hamiltonian is ($\hbar=1$)
\begin{equation}
\hat{H}=\omega_{0} \hat{o}^{\dagger}\hat{o} + \sum_{k} g_k (\hat{o}^{\dagger} \hat{a}_{k}+ \hat{a}^{\dagger}_{k} \hat{o})+\sum_{k} \omega_k \hat{a}^{\dagger}_{k}\hat{a}_{k},
\label{eq:Hamiltonian}
\end{equation}
where $\hat{o}$ and $\hat{a}_{k}$ are the annihilation operator of the system with frequency $\omega_{0}$ and the $k$-th environmental mode with frequency $\omega_{k}$, and $g_k$ is the coupling strength. The coupling can be further characterized by the spectral density $J(\omega)=\sum_{k} |g_{k}|^{2} \delta(\omega-\omega_{k})$ \cite{Vega2017}. The operator $\hat{o}=\hat{\sigma}_-$ with $\hat{\sigma}_-=|g\rangle\langle e|$ is the transition operator between the two levels for the discrete-variable system, $\hat{o}=\hat{b}$ is the annihilation operator for the bosonic continuous-variable system. One can check that  the total excitation number $\hat{N}=\hat{\sigma}_+\hat{\sigma}_-+\sum_k \hat{a}^{\dagger}_{k}\hat{a}_{k}$ for both of the systems is conserved. Thus, the Hilbert space is split into the subspaces with a definite excitation number $N$.

The dynamics of the open system is obtainable by tracing out the environmental degrees of freedom from $\rho_\text{T}(t)$ satisfying the Liouville
equation $\dot{\rho}_\text{T}(t)=-i[\hat{H},\rho_\text{T}(t)]$. However, it is difficult to be exactly done. A widely used reduction is the Markovian approximation \cite{Kossakowski1972,CrispinGardiner2004,Ackerhalt1973}, which works under the conditions that the system-environment coupling is weak and the environmental correlation time is much smaller than the one of the systems. Many systems, such as quantum dot \cite{Khaetskii2002,Merkulov2002} and superconducting qubit \cite{Ithier2005}, have shown the conditions where this approximation is inapplicable. It inspires particular attention to the non-Markovian dynamics \cite{Breuer2016,Vega2017}.

Supposing that the coupling is turned on at the initial time, we have $\rho_\text{T}(0)=\rho(0) \otimes \rho_\text{E}$, where $\rho(0)$ and $\rho_\text{E}$ are the initial state of system and environment, respectively. We consider that the environment is initially in a vacuum state. The dynamics governed by Eq. \eqref{eq:Hamiltonian} in this case for the discrete-variable system is exactly solvable because only the Hilbert subspaces with $N=0$ and $1$ are involved \cite{Tong2010}. The exact dynamics for the continuous-variable system is obtainable by the Feynman-Vernon influence functional theory \cite{FEYNMAN1963118} in the coherent-state representation \cite{An2007}. After tracing out the environmental degrees of freedom, we obtain the non-Markovian master equation for both of the systems as
\begin{equation}
\dot{\rho}(t)=-i\omega(t)[\hat{o}^\dag\hat{o},\rho(t)]+{\gamma(t)\over 2}[2\hat{o}\rho(t)\hat{o}^\dag-\{\hat{o}^\dag\hat{o},\rho(t)\}],\label{masteqd}
\end{equation}
where $\gamma(t)-i\omega(t)=\dot{u}(t)/u(t)$ with $u(t)$ satisfying
\begin{equation}
\dot{u}(t)+i\omega_0u(t)+\int_0^t f(t-\tau)u(\tau)d\tau=0,\label{eq:evolution-CF}
\end{equation}under $u(0)=1$. Here the kernel function is $f(t-\tau)=\int_0^\infty J(\omega)e^{-i\omega(t-\tau)}d\omega$.

A Laplace transform can linearize Eq.~\eqref{eq:evolution-CF} into $\tilde{u}(z)=[z+i\omega_0+\int_0^\infty{J(\omega)\over z+i\omega}d\omega]^{-1}$. The solution of $u(t)$ is obtained by the inverse Laplace transform of $\tilde{u}(z)$, which can be done by finding its pole from
\begin{equation}
y(E)\equiv\omega_0-\int_0^\infty{J(\omega)\over\omega-E}d\omega =E,~(E=iz).\label{eigen}
\end{equation}
Note that the roots $E$ of Eq. \eqref{eigen} is just the eigenenergies of the total system \eqref{eq:Hamiltonian} in the single-excitation space. Specifically, expanding the eigenstate as $|\Psi\rangle=(x\hat{o}^{\dagger}+\sum_{k}y_{k}\hat{b}_{k}^{\dagger})|0,\{0_k\}\rangle$ and substituting it into $\hat{H}|\Psi\rangle=E|\Psi\rangle$ with $E$ being the eigenenergy, we have $(E-\omega_{0})x=\sum_{k}g_{k}y_{k}$ and $y_{k}=g_{k}x/(E-\omega_{k})$. They readily lead to Eq. \eqref{eigen}. Since $y(E)$ is a decreasing function in the regime $E < 0$, Eq. \eqref{eigen} has one isolated root $E_b$ in this regime provided $y(0) < 0$. While $y(E)$ is ill defined when $E>0$, Eq. \eqref{eigen} has infinite roots in this regime forming a continuous energy band. We call the eigenstate of the isolated eigenenergy $E_b$ bound
state \cite{Tong2010,Bai2019}. After the inverse Laplace transform, we obtain
\begin{equation}
u(t)=Ze^{-iE_b t}+\int_{0}^{\infty}\frac{J(E)e^{-iE t}dE}{[E-\omega_{0}-\Delta(E)]^{2}+[\pi J(E)]^2},\label{eq:eq13}
\end{equation}
where $\Delta(E)=\mathcal{P} \int_0^\infty \frac{J(\omega)}{E-\omega}d\omega $ with $\mathcal{P}$ being the Cauchy principal value. The first term with $Z=[1+\int_0^\infty{J(\omega)d\omega\over(E_b-\omega)^2}]^{-1}$ is from the bound state, and the second one is from the band energies. Oscillating with time in continuously changing frequencies, the integral tends to zero in the long-time condition due to out-of-phase interference. Thus, if the bound state is absent, then $\lim_{t\rightarrow\infty} u(t)= 0$ characterizes a complete dissipation, while if the bound state is formed, then $\lim_{t\rightarrow\infty} u(t)=Ze^{-iE_b t}$ implies a dissipation suppression. A widely used type of spectral density reads $J(\omega)=\eta\omega^{s}\omega_{c}^{1-s} e^{-\omega/\omega_{c}}$, where $\eta$ is a dimensionless coupling constant, $\omega_{c}$ is a cutoff frequency, and $s$ is the Ohmicity parameter. Depending on the value of $s$, the environments are classified into a sub-Ohmic one ($0<s<1$), an Ohmic one ($s=1$), and a super-Ohmic one ($s>1$) \cite{Leggett1987}. We can check that the bound state is formed for the Ohmic-family spectral density when $\omega_0<\eta\omega_c\underline{\Gamma}(s)$, where $\underline{\Gamma}(s)$ is the Euler's gamma function.

From the above analysis, we see that the non-Markovian dissipation dynamics for both of the discrete- and continuous-variable systems intrinsically depends on the character of the energy spectrum of the total system. In the following, we review the profound impact of the bound state on the different nonequilibrium physics of the systems.

\subsection{Entanglement preservation }\label{sec:decoh-EM}

Quantum entanglement describes the nonclassical correlations between quantum subsystems, which was first recognized by Einstein, Podolsky, and Rosen \cite{Einstein1935}. It is an important resource to implement quantum protocols \cite{Horodecki2009}, e.g. quantum communication \cite{Bennett2014,Ekert1991,Gisin2002,Bennett1992,Mattle1996}, quantum computation  \cite{Feynman1982,Ekert1996,Law2001}, and quantum metrology \cite{Giovannetti2004,Leibfried2004,Giovannetti2011,Escher2011}. However, it is generally degraded by the ubiquitous decoherence in the microscopic world. It was found that, quite different from the asymptotically decaying to zero of quantum coherence of a single two-level system, the entanglement of a bipartite two-level system influenced by two independent Markovian dissipative environments abruptly disappears in a finite time duration \cite{Yu2004,Almeida2007,Yu2009}. It is called entanglement sudden death. Further studies revealed that the dead entanglement can partially revive due to the non-Markovian effect \cite{Bellomo2008,Maniscalco2008,Xu2010}. Such finite extension of the entangled time is obviously not enough for quantum protocols and one generally desires to preserve the entanglement in the long-time limit. Indeed, it
was shown that some entanglement can be preserved in the steady state by engineering a photonic crystal structured environment \cite{Bellomo2008a}. Actually, as demonstrated in the following, the entanglement preservation is a universal consequence of the existence of the system-environment bound state and the non-Markovian effect. The bound state provides the ability and the non-Markovian effect provides the dynamical way to preserve the entanglement \cite{Tong2010,Tong2010a,Lue2013}.

Consider two independent two-level systems interacting with their respective dissipative environment. The dynamics of each two-level system is governed by Eq. \eqref{masteqd} with $\hat{o}=\hat{\sigma}_-$ \cite{Tong2010a}. Solving the master equation under the initial condition $|\psi(0)\rangle=\frac{1}{\sqrt{2}}(|gg\rangle +|ee\rangle)$, we obtain
\begin{eqnarray}
\rho(t)=\frac{1}{2}\big\{\big{[}|u(t)|^{2}|e\rangle\langle e|+(1-|u(t)|^{2})|g\rangle\langle g|\big{]}^{\otimes 2}+|g\rangle\langle g|^{\otimes 2}+[u(t)^{2}|e\rangle\langle g|^{\otimes 2}+\mathrm{H}.\mathrm{c}.]\big\}.\label{nrhot}
\end{eqnarray} The entanglement of a bipartite two-level system is quantified by concurrence \cite{Wootters1998}. It is defined as $C(\rho)=\text{max}\{0,\lambda_1-\lambda_2-\lambda_3-\lambda_4 \}$, where $\lambda_i$ is the $i$-th decreasing ordered eigenvalues of $R\equiv\sqrt{\sqrt{\rho}\tilde{\rho}\sqrt{\rho}}$ and $\tilde{\rho}$ is defined as $\tilde{\rho}\equiv (\hat{\sigma}_y\otimes\hat{\sigma}_y)\rho^*(\hat{\sigma}_y\otimes\hat{\sigma}_y)$ \cite{Hildebrand2007,Horodecki2009}. Then we can obtain that the concurrence of Eq. \eqref{nrhot} is $C(t)=\text{max}\{0,|u(t)|^4\}$. Combining with the knowledge that $u(t)$ is preserved when the bound state exists, we find that the formation of the bound state can also preserve the entanglement, which paves the way to stably protect entanglement from dissipation.

\begin{figure}
	\centering
    \includegraphics[width=0.8\textwidth]{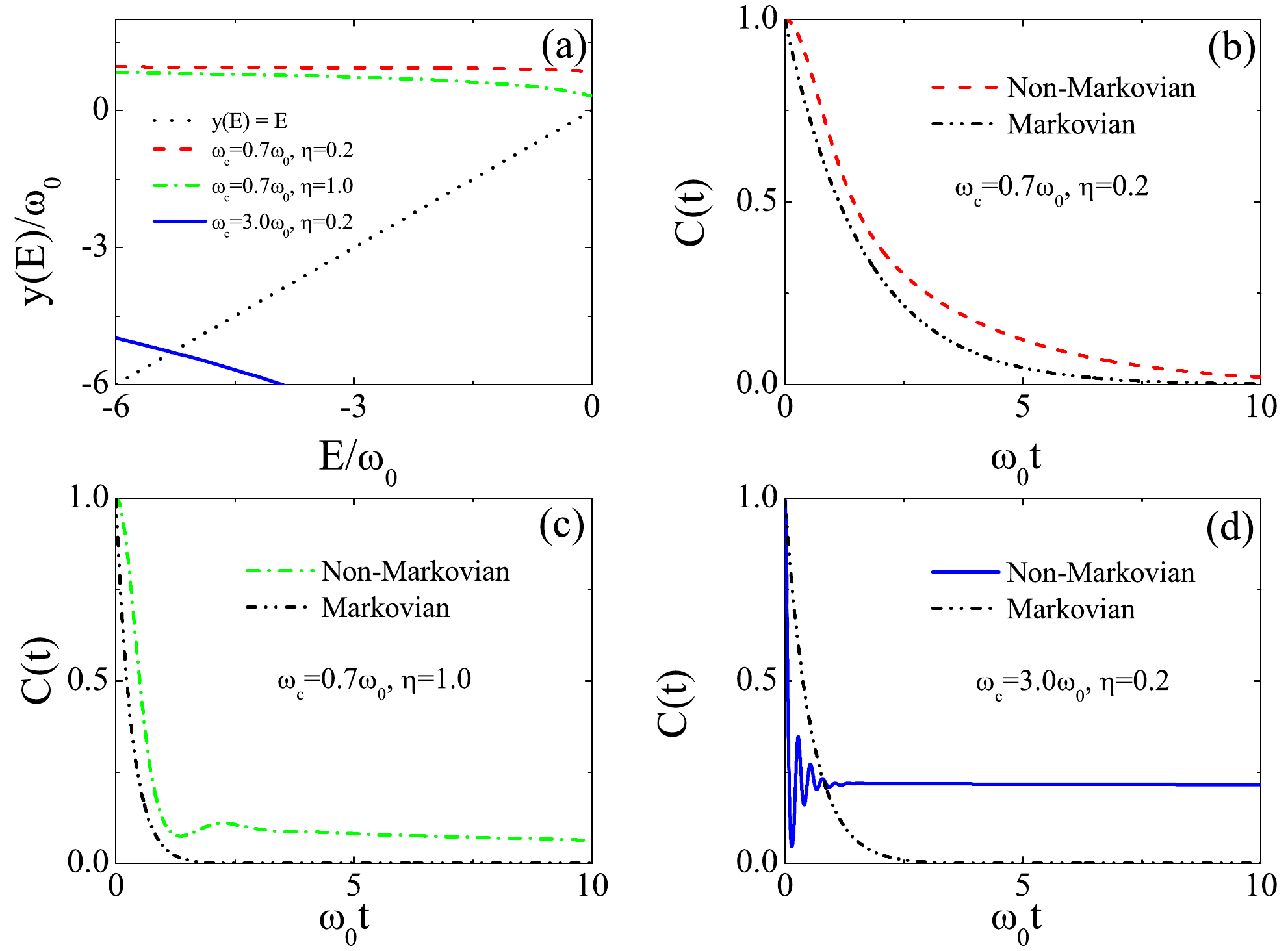}
    \caption{(a) Diagrammatic solution of Eq. \eqref{eigen} with different parameters of a super-Ohmic spectral density $s=3$. Entanglement evolution with parameters in the absence (b, c) and presence (d) of the bound state. Reproduced figures from \cite{Tong2010}.}
    \label{fig:EM_fig1}
\end{figure}

Taking the super-Ohmic spectral density $J(\omega)=\eta\omega^{s}\omega_{0}^{1-s} e^{-\omega/\omega_{c}}$ with $s=3$, where the bound state is formed when  $\omega_{0}-\eta\omega_0(\omega_{c}/\omega_0)^s\underline{\Gamma}(s)<0$, as an example \cite{Tong2010}, Fig. \ref{fig:EM_fig1}(a) shows the diagrammatic solution of Eq. \eqref{eigen} for different parameter values. It reveals that as long as $\omega_0<2\eta\omega_0(\omega_c/\omega_0)^3$, the bound state is formed. The corresponding entanglement dynamics are shown in Figs. \ref{fig:EM_fig1}(b), \ref{fig:EM_fig1}(c), and \ref{fig:EM_fig1}(d). Under the Markovian approximation, the concurrence always decays to zero irrespective of the existence of the bound state, whereas the exact non-Markovian dynamics reveals that the entanglement is preserved when the bound state is formed. The one-to-one correspondence between the formation of bound state and entanglement preservation clearly signifies the ability of the bound state in protecting the entanglement from decoherence.

Note that the entanglement preservation mechanism given above is also applicable to other spectral density \cite{Tong2010} and other systems, e.g., nitrogen-vacancy centers embedded into the planar photonic-crystal cavities \cite{Yang2013}, continuous-variable system \cite{An2008,Lue2013}, and quantum emitters coupled to the localized surface plasmons \cite{Yang2019}.

\subsection{Quantum speed limit}\label{sec:decoh-QS}

The Heisenberg's uncertainty principle sets a fundamental limit to measurement precision. One well-known example is that the precision of position is limited by the uncertainty of momentum as $\Delta x \ge \hbar/ (2\Delta p)$. Such limitation also exists for the detection of time, which is known as the quantum speed limit (QSL) \cite{Anandan1990, Mandelstam1991,Vaidman1992,Lloyd2000}. It sets a bound on the minimal time a system needs to evolve between two distinguishable states \cite{Luo2004,Margolus1998}. The QSL time for isolated systems is determined by the maximum of the Mandelstam-Tamn bound $\tau_\text{ML}=\pi \hbar /(2 \Delta E)$ \cite{ Mandelstam1991,Vaidman1992} and Margolus-Levitin bound $\tau_\text{ML}=\pi \hbar/(2 \bar{E})$ \cite{Margolus1998}, where $\Delta E$ and $\bar{E}$ are the variation and the average of the Hamiltonian over the initial state, respectively. The QSL time for open systems characterizes the most efficient response of the system to the environmental influences. The Mandelstam-Tamm-type bound is generalized to open systems by using positive non-unitary maps \cite{Taddei2013, Campo2013}. Based on a geometric approach, a unified bound including both Mandelstam-Tamm and Margolus-Levitin types has been formulated \cite{Pires2016}. The QSL has attracted considerable attention because the efficient speedup of quantum systems plays remarkable roles in various areas of quantum physics, including nonequilibrium thermodynamics \cite{Deffner2010}, quantum metrology \cite{Alipour2014}, quantum optimal control \cite{Gajdacz2015}, quantum computation \cite{Tran2020}, and quantum communication \cite{Yung2006}.

The previous studies revealed that the non-Markovian effect characterized by non-Markovianity can speed up the evolution of open systems \cite{Deffner2013,Cimmarusti2015}. As a dynamical quantity, the non-Markovianity could not be seen as an essential reflection on the QSL of open systems. Furthermore, such attribution of the quantum speedup to the non-Markovianity is experimentally meaningless because the non-Markovianity is not an observable. Actually, it is the formation of the system-environment bound state that is the essential reason for the quantum speedup.

\begin{figure}
	\centering
	\includegraphics[width=0.65\textwidth]{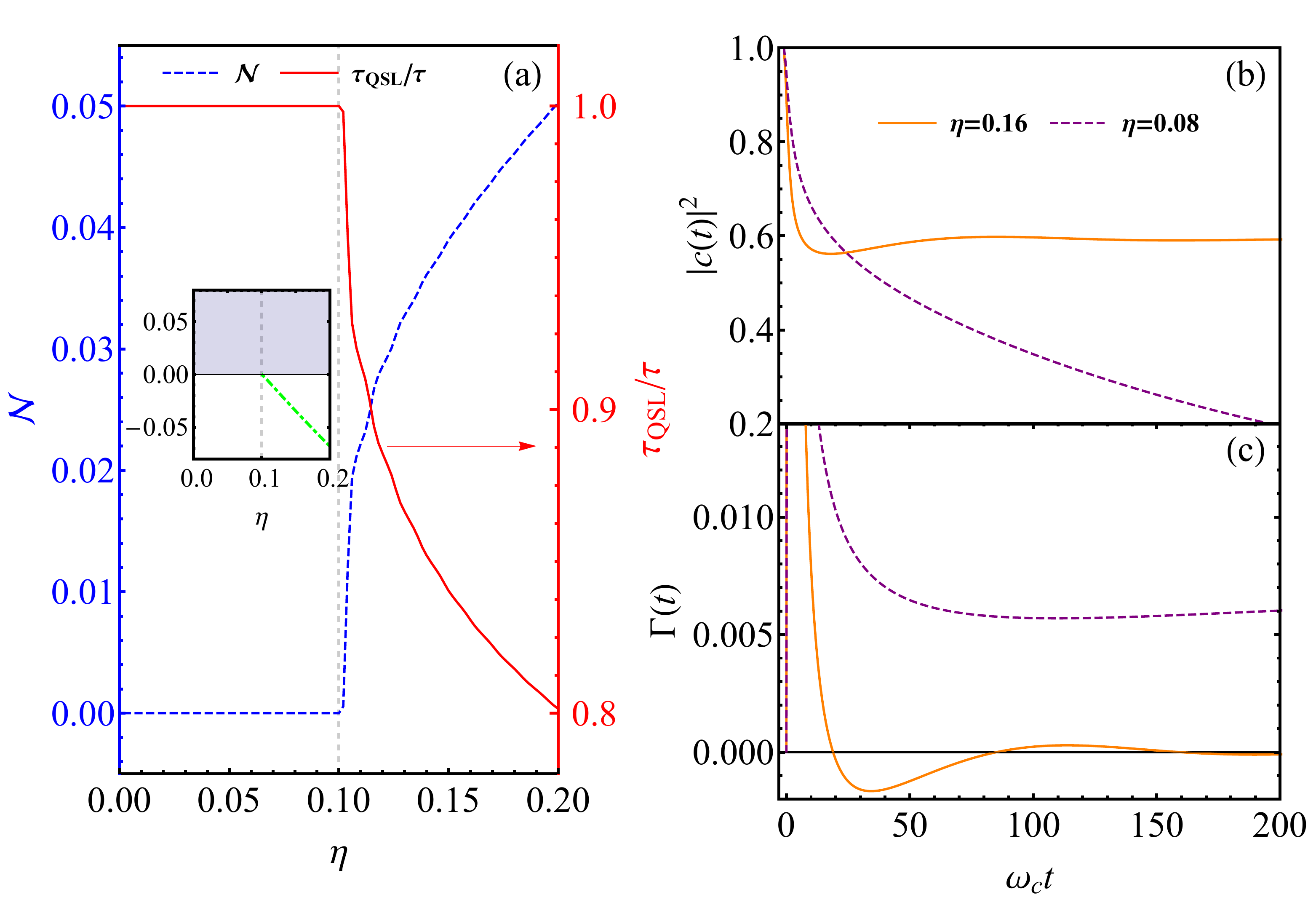}
	\includegraphics[width=0.3\textwidth]{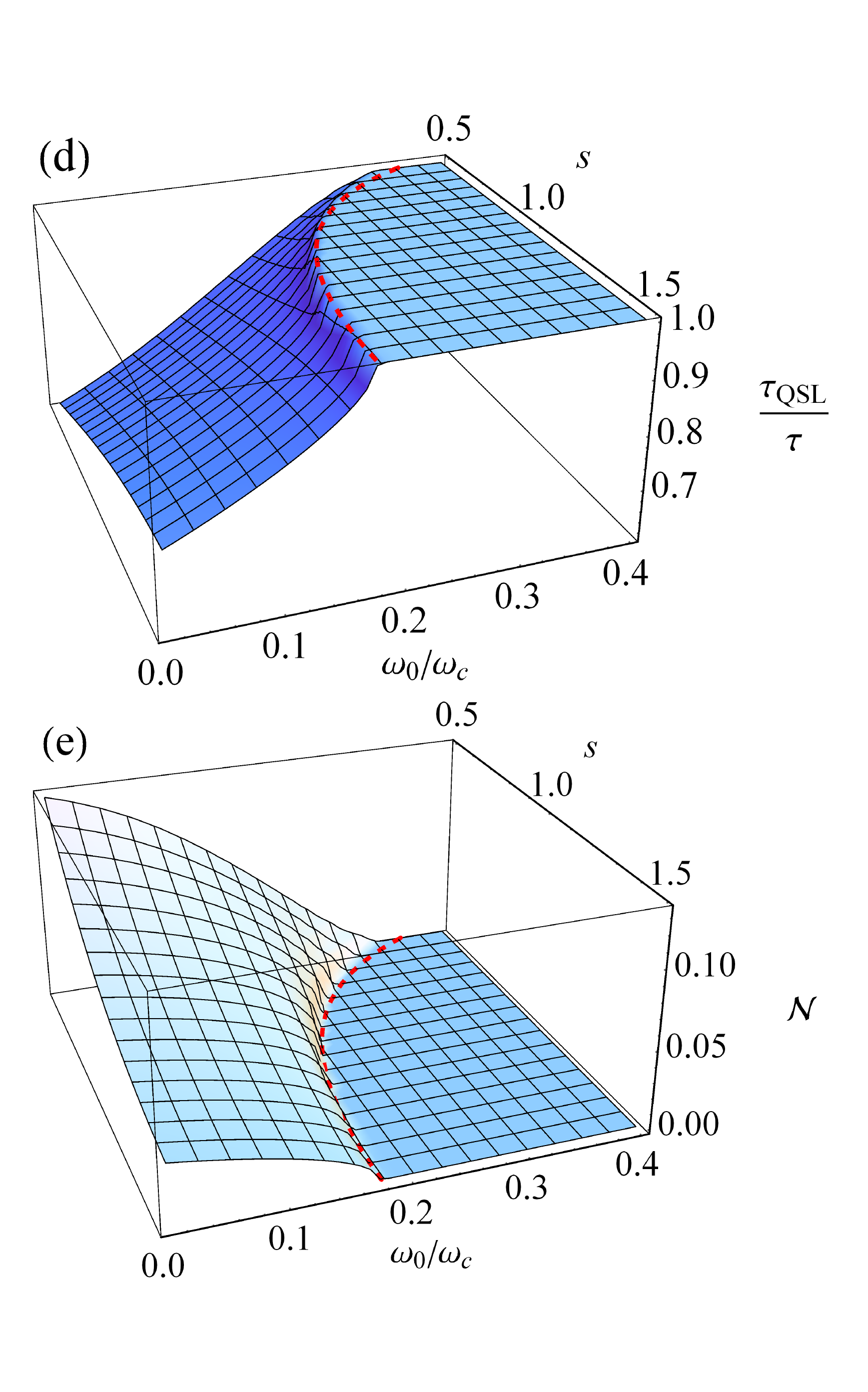}
	\caption{(a) Non-Markovianity $\mathcal{N}$ (blue dashed line) and QSL time $\tau_\text{QSL}$ (read solid line) for the Ohmic spectral density as a function of coupling constant $\eta$. The inset is the energy spectrum of the total system, where green dot-dashed line denotes the bound state. The obtained $|u(t)|^2$ (b) and the corresponding decay rate $\gamma(t)$ (c) with and without bound state. Parameters in (a), (b), and (c) are $\omega_{c}\tau=800$, $\omega_0/\omega_{c}=0.1$, and $s=1$, which determines that the bound state is formed when $\eta>0.1$. QSL time $\tau_\text{QSL}$ (d) and non-Markovianity $\mathcal{N}$ (e) as a function of the spectral power index $s$ and the system frequency $\omega_0$ for the spectral density $J(\omega)=\eta\omega^s\omega_c^{1-s}e^{-\omega/\omega_c}$. The red dashed line shows the critical values for forming the bound state. Parameters in (d) and (e) are $\omega_c\tau = 800$ and $\eta = 0.2$. Reproduced figures from \cite{Liu2016}. }
	\label{fig:QSL-BS}
\end{figure}
To demonstrate this result, we consider again a dissipative two-level system \cite{Liu2016}. Its dynamics follows the master equation \eqref{masteqd} with $\hat{o}=\hat{\sigma}_-$. The QSL time between an initial state $\rho (0)=|\psi _{0}\rangle \langle \psi _{0}|$ and its target state $\rho(\tau)$ is defined as $\tau _\text{QSL}=\sin^{2} \mathcal{B}[\rho (0),\rho (\tau )]/\Lambda_\tau^{(\infty)}$ \cite{Deffner2013}, where $\mathcal{B}[ \rho (0),\rho (\tau )] \equiv\arccos \sqrt{\langle\psi _{0}\vert \rho (\tau )\vert \psi _{0}\rangle }$ is the Bures angle between $\rho (0) $ and $\rho (\tau )$, and $\Lambda_{\tau }^{(\infty)}=( 1/\tau )\int_{0}^{\tau }dt||\dot{\rho} (t)|| _{\infty}$ with $\Vert \hat{A}\Vert_{\infty}$ equaling to the largest eigenvalue of $\sqrt{\hat{A}^\dag \hat{A}}$. From Eq. \eqref{masteqd}, the QSL time for this system
\begin{equation}
\tau _\text{QSL}=\frac{1-|u(\tau)|^{2}}{(1/\tau )\int_{0}^{\tau }| \partial _{t}|u(t)|^{2}| dt},  \label{eq:QSL}
\end{equation} is achieved when $|\psi_0\rangle=|e\rangle$. In order to make a comparison with the previous works, we also calculate the non-Markovianity defined as $\mathcal{N}=\max_{\rho _{1,2}(0)}\int_{\sigma >0}dt\sigma (t,\rho _{1,2}(0))$, where $\sigma (t,\rho_{1,2}(0))=\dot{D}(\rho _{1}(t),\rho _{2}(t))$ is the change rate of the trace distance $D(\rho _{1}(t),\rho _{2}(t))=\frac{1}{2}\text{Tr}\vert \rho _{1}(t)-\rho_{2}(t)\vert $ between states $\rho_{1,2}(t)$ evolving from their respective initial forms $\rho_{1,2}(0)$ \cite{Breuer2009}. For the two-level system, it has been proven that the optimal pair of initial states to maximize $\mathcal{N}$ are $\rho_{1}(0)=|g\rangle\langle g|$ and $\rho_{2}(0)=|e\rangle\langle e|$ \cite{Deffner2013}, which means $D(\rho _{1}(t),\rho _{2}(t))=|u(t)|^2$. Then one can readily verify $\mathcal{N}={1\over 2}\big[|u(\tau)|^2-1+\int_0^\tau |\partial_t|u(t)|^2|dt\big]$, which connects to $\tau_\text{QSL}$ as $\tau_\text{QSL}=\tau{1-|u(\tau)|^2\over 1-|u(\tau)|^2+2\mathcal{N} }$.

The numerical results for the Ohmic-family spectral density are shown in Fig. \ref{fig:QSL-BS}. From Fig. \ref{fig:QSL-BS}(a) one can find that both the non-Markovianity and quantum speedup show a perfect correspondence with the formation of bound state. In the absence of the bound state, there is no non-Makrovianity and quantum speedup. Whereas, an abrupt increase of the non-Markovianity and an abrupt decrease of the QSL time are observed when the bound state is formed. The corresponding excited-state population $\text{Tr}[\hat{\sigma}_+\hat{\sigma}_-\rho(t)]=|u(t)|^2$ and the decay rate $\gamma(t)$ are plotted in Figs. \ref{fig:QSL-BS}(b) and \ref{fig:QSL-BS}(c), respectively. We can see that $\gamma(t)$ in the absence of the bound state tends to a positive constant after a short-time jolt. Its complete positivity causes $|u(t)|^2$ to decay to zero monotonically. Since the system equilibrates in an asymptotic manner to its ground state, there is no more efficient evolution than the evolution characterized by $\tau$ and thus $\tau_\text{QSL} =\tau$. When the bound state is formed, the competition between the environmental backaction and the dissipation causes $\gamma(t)$ to transiently take a negative value and asymptotically approaches zero. Thus, after some short-time oscillations, $|u(t)|^2$ tends
to a finite value. The transient increase of $|u(t)|^2$ causes the increase of the distinguishability and the decrease of $\tau_\text{QSL}$. The system here has a great capacity to speedup. Hence the bound state supplies a latent capability, while the non-Markovianity only supplies the dynamical way to the system for quantum speedup. The result is confirmed in the general cases with changing $\omega_{0}$ and $s$ in Figs. \ref{fig:QSL-BS} (d) and \ref{fig:QSL-BS}(e).

The relationship between the bound state and quantum speedup reveals a mechanism for quantum speedup. It may be of both theoretical and experimental interests in exploring the ultimate QSL in realistic environments, and may open new perspectives for devising active quantum speedup devices.

\subsection{Noncanonical thermalization}\label{sec:decoh-CT}

In statistical mechanics, systems in equilibrium with its weakly coupled environment are described by a canonical state at the same temperature as the environment. Being called canonical thermalization, this is based on the assumption of ergodicity. The assumption relies on the subjective lack of knowledge of systems. Therefore, how and when systems microscopically equilibrate from any initial state to a unique canonical state has attracted much attention \cite{Popescu2006, Linden2009, Tasaki1998, Goldstein2006, Reimann2010, Lychkovskiy2010, Lee2012, Genway2013, Rigol2008, Ponomarev2011, Banuls2011, Polkovnikov2011}. It was revealed that the assumption of ergodicity could be abandoned by examining the entanglement induced by the system-environment interaction, by which the canonical state can be achieved without referring to ensembles or time averages \cite{Popescu2006}. It inspires the studies of canonical thermalization from the dynamics of open quantum systems \cite{Geva2000, Mori2008, Subasi2012, Pagel2013, Rancon2013}. It is expected that equilibrium statistical physics should be an emergent steady-state limit of the nonequilibrium dynamics. It was shown that the canonical thermalization is valid only in the limit of the vanishing coupling and the memoryless correlation function of the environment \cite{Geva2000,Subasi2012}. However, in real physical situations, e.g. photonic crystal \cite{Hoeppe2012} and semiconductor phonon \cite{Tahara2011} environments, low-dimensional solid-state system \cite{Galland2008}, and optical systems \cite{Liu2011,Madsen2011}, these conditions are generally invalid and the strong memory effect of the environment may even cause the anomalous decoherence behavior \cite{Liu2013, Yang2017, Yang2019}. Therefore, to microscopically bridge between an arbitrary initial state of the open system and its equilibrium state via studying its equilibration dynamics is desired to check the validity regime of the assumption of ergodicity \cite{Yang2014}.

\begin{figure}[tpb]
  \centering
\includegraphics[width=0.9\textwidth]{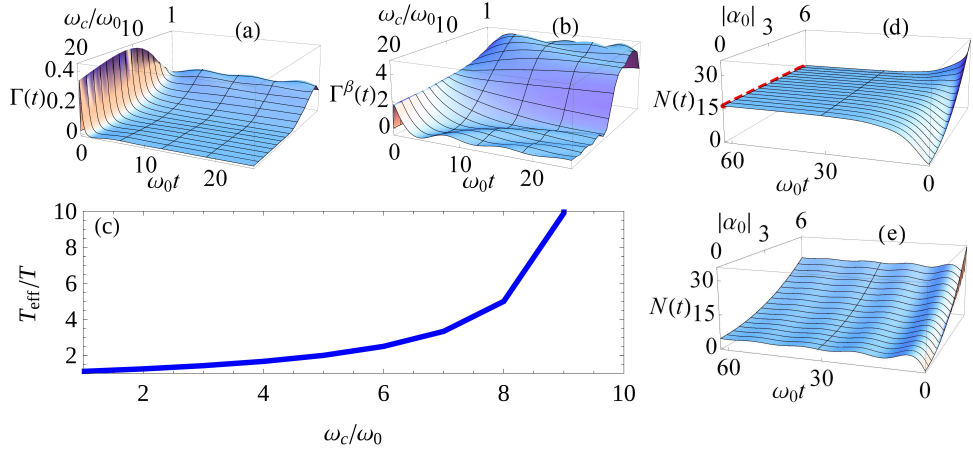}\\
\caption{$\Gamma(t)$ (a) and $\Gamma^\beta(t)$ (b) in different $\omega_c$. (c) $T_\text{eff}$ evaluated when $t=100/\omega_0$. $\eta=0.1$, $\beta=0.1/\omega_0$, and $s=1$ have been used. Mean photon number $N(t)$ in different initial condition $\alpha_0$ when $\eta=0.05$ (d) and $0.2$ (e), where the red dashed line is obtained from Eq. \eqref{cano}. Reproduced figure from \cite{Yang2014}.}
  \label{fig:Eff-T}
\end{figure}

To address this issue, we consider a continuous-variable system described by a quantum harmonic oscillator interacting with a finite-temperature environment \cite{Yang2014}. The Hamiltonian follows Eq. \eqref{eq:Hamiltonian} with $\hat{o}=\hat{b}$. After generalizing Eq. \eqref{masteqd} to the finite-temperature case, we obtain the exact non-Markovian master equation as
\begin{equation}
\begin{split}
\dot{\rho}(t)=&-i\Omega(t)\big{[}\hat{b}^{\dagger}\hat{b},\rho(t)\big{]}+\Big{[}\Gamma(t)+\frac{\Gamma_{\beta}(t)}{2}\Big{]}\big{[}2\hat{b}\rho(t)\hat{b}^{\dagger}-\{\hat{b}^{\dagger}\hat{b},\rho(t)\}\big{]}\\
&+\frac{\Gamma_{\beta}(t)}{2}\big{[}2\hat{b}^{\dagger}\rho(t)\hat{b}-\{\hat{b}\hat{b}^{\dagger},\rho(t)\}\big{]},\label{fntmastq}
\end{split}
\end{equation}
where $\Gamma(t)+i\Omega(t)=-\dot{u}(t)/u(t)$ and $\Gamma_{\beta}(t)=\dot{v}(t)+2v(t)\Gamma(t)$. The function $u(t)$ satisfies Eq. \eqref{eq:evolution-CF} and $v(t)=\int_{0}^{t}dt_{1}\int_{0}^{t}dt_{2}u^{*}(t_{1})\mu(t_{1}-t_{2})u(t_{2})$, where $\mu(x)=\int_{0}^{\infty}\bar{n}(\omega)J(\omega)e^{-i\omega x}d\omega$ with $\beta=(k_BT)^{-1}$ and $\bar{n}(\omega)=(e^ {\beta \omega }-1)^{-1}$. According to the detailed balance condition, as long as $\Gamma(t)$ and $\Gamma^\beta(t)$ are positive, the steady state of Eq. \eqref{fntmastq} must be
\begin{equation}
\rho(\infty)=\sum_{n=0}^{\infty }\frac{%
[\Gamma^\beta(\infty )/(2\Gamma(\infty ))]^{n}}{[1+\Gamma^\beta(\infty )/(2\Gamma(\infty ))]^{n+1}}|n\rangle \langle n|,\label{rhoinf}
\end{equation}
which defines a unique canonical state irrespective of the initial state
\begin{eqnarray}
\rho _\text{con}\equiv \rho(\infty)=e^{-\beta_\text{eff}\hat{H}_\text{s}}/ \text{Tr}_\text{s}e^{-\beta_\text{eff}\hat{H}_\text{s}},~(\beta_\text{eff}=1/k_B T_\text{eff}),\label{cano}
\end{eqnarray} with $T_\text{eff}=\frac{\omega_0 }{ k_B}\{\ln [1+\frac{2\Gamma (\infty)}{\Gamma ^{\beta }(\infty)}]\}^{-1}$.
The effective temperature $T_\text{eff}$ reduces to the environmental temperature $T$ in the Markovian limit. This can be verified from the Markovian solution of Eq. \eqref{eq:evolution-CF}, i.e. $u_\text{MA}(t)=e^{-(\Gamma_0+i\Omega_0)t }$ with $\Gamma_0=\pi J(\omega_0)$ and $\Omega_0=\omega_0+\mathcal{P}\int{J(\omega)\over \omega_0-\omega}d\omega$, which leads to $\Gamma_\text{MA}(t)=\kappa$, $\Omega_\text{MA}(t)=\Omega_0$, and $\Gamma^\beta_\text{MA}(t)=2\kappa \bar{n}(\omega_0)$. We readily have $T_\text{eff,MA}=T$. Thus, the canonical ensemble assumption in statistical mechanics can be dynamically confirmed under the Markovian approximation.

In the non-Markovian situation, the obtained $\Gamma(t)$, $\Gamma^\beta(t)$, and $T_\text{eff}$ for the Ohmic spectral density are plotted in Fig. \ref{fig:Eff-T}. In the absence of the bound state when $\omega_c< 10\omega_0$, both $\Gamma(t)$ and $\Gamma^\beta(t)$ tend to positive constant values asymptotically. Their positivity guarantees the validity of Eq. \eqref{rhoinf}. Consequently, the system approaches a canonical state governed by Eq. (\ref{cano}), as shown in Fig. \ref{fig:Eff-T}(c), irrespective of in what state the system initially resides. Here it is qualitatively consistent with the one under the Markovian approximation. However, whenever the bound state is formed in the regime $\omega_c\ge 10\omega_0$ of Fig. \ref{fig:Eff-T}, $\Gamma(t)$ and $\Gamma^\beta(t)$ possess transient negative values and approach zero asymptotically. Their vanishing values cause the ill definition of the canonical state (\ref{cano}). In Fig. \ref{fig:Eff-T}(c), we really see that $T_\text{eff}$ changes to be divergent when the parameters tend to the critical point forming the bound state. Another interesting observation is that even when the equilibration to a canonical state is valid in the absence of the bound state, $T_\text{eff}$ still shows a dramatically quantitative difference from $T$. Figure \ref{fig:Eff-T}(d) indicates that the equilibrium state matching with Eq. \eqref{rhoinf} is independent of the initial condition when the bound state is absent. However, this is not true as long as the bound state is formed [see Fig. \ref{fig:Eff-T}(e)]. This result confirms again the breakdown of canonical thermalization due to the formation of the bound state.

Building a unified framework to treat the issues in equilibrium and nonequilibrium statistical mechanics, the result gives a microscopic description of the canonical thermalization and reveals the working condition of the assumption of ergodicity as the basis of the equilibrium statistical mechanics.

\subsection{Quantum metrology}\label{sec:decoh-QM}

Quantum metrology employs quantum resources, such as entanglement and squeezing, to attain a measurement precision surpassing the limit achievable in classical physics \cite{Giovannetti2004, Giovannetti2006,  Braunstein1994, Degen2017, Pezze2018}. It has been found that the metrology precision with a $N$-body entangled state reaches the Heisenberg limit (HL) scaling $N^{-1}$, which is far better than the shot-noise limit (SNL) $N^{-1/2}$ revealed by $N$-body product state \cite{Giovannetti2004}. Such an enhancement inspires the application in gravitation wave detection \cite{Tse2019, Acernese2019} and atomic clocks \cite{Ludlow2015, Kruse2016}. The main obstacle of quantum metrology in practical applications is decoherence. It has been shown that the Markovian decoherence degrades the scaling from the HL back to the SNL \cite{Huelga1997, Escher2011}. Some studies revealed that the non-Markovian effect can retrieve quantum superiority to beat the SNL \cite{Matsuzaki2011, Chin2012, Macieszczak2015}. However, it works only in the short encoding time condition. The precision gets worse in the long-time condition. Therefore, how to achieve metrology precision surpassing the SNL in the long encoding time condition in the presence of the dissipative environment is of significance in the practical application of quantum metrology.

\subsubsection{Quantum metrology based on Mech-Zehnder interferometer}
To estimate a frequency $\gamma$ of a system, we prepare two modes of optical fields with frequency $\omega_0$ and the initial state $|\Psi_\text{in}\rangle$ as the probe to input into the Mech-Zehnder interferometer (MZI). The input-output relation of the MZI reads
$|\Psi_\text{out}\rangle=\hat{V}\hat{U}_0(\gamma,t)\hat{V}|\Psi_\text{in}\rangle$, where $\hat{V}=\exp[i\frac{\pi}{4}(\hat{b}_{1}^{\dagger}\hat{b}_{2}+\hat{b}_{2}^{\dagger}\hat{b}_{1})]$ is the action of $\text{BS}_i$ and $\hat{U}_0(\gamma,t)=\exp(-{i\hat{H}_0t/\hbar})$ with $\hat{H}_0=\hbar\omega_0\sum_{m=1,2}\hat{b}_m^\dag\hat{b}_m+\hbar\gamma\hat{b}_2^\dag\hat{b}_2$ describes the free evolution of the probe and its coupling to the system for encoding. The photon difference $\hat{M}=\hat{b}^\dag_1\hat{b}_1-\hat{b}^\dag_2\hat{b}_2$ is measured by $\text{D}_i$. If we use the squeezed state $|\Psi_\text{in}\rangle=\hat{D}_{\hat{b}_{1}} \hat{S}_{\hat{b}_{2}} |0,0\rangle$, where $\hat{D}_{\hat{b}}=e^{\alpha \hat{b}^{\dagger}-\alpha^{\ast} \hat{b}}$ with $\alpha = |\alpha| e^{i \varphi}$, $\hat{S}_{\hat{b}}=e^{\frac{1}{2} (\xi^{\ast} \hat{b}^{2}-\xi \hat{b}^{\dagger 2}) }$ with $\xi=r e^{i \phi}$, and $|0,0\rangle$ is the vacuum state, then we have the best precision of estimating $\gamma$ via $\delta \gamma=\frac{\delta M}{|\partial\bar{M}/\partial\gamma|}$ as \cite{Caves1981}
\begin{equation}
\min\delta \gamma =\frac{[(1-\beta)e^{-2r}+\beta]^{1\over2}}{t\sqrt{N}|1-2\beta|},~(\beta\equiv\sinh^2r/N)
\end{equation}
when $\phi =2\varphi $ and $\gamma t={(2m+1)\pi /2}$ for $m\in \mathbb{Z}$. If the squeezing is absent, then $\min\delta \gamma|_{\beta=0}=(t N_0^{1/2})^{-1}$ with $N_0=|\alpha|^2$ is just the SNL. For $\beta\neq 0$, using $e^{-2r}\simeq 1/(4\sinh^2r)$ for $N\gg1$ and optimizing $\beta$, we obtain $\min\delta\gamma|_{\beta=(2\sqrt{N})^{-1}}=(tN^{3/4})^{-1}$, which is called Zeno limit. It beats the SNL and manifests the superiority of the squeezing in metrology.

In practice, the above ideal encoding dynamics is obscured by the decoherence. The decoherence in MZI comes from the photon loss. Consider that the second optical field is affected by photon loss during the parameter encoding process, then its dynamics is governed by Eq. \eqref{masteqd} with $\hat{o}=\hat{b}_2$. We can calculate \cite{Bai2019}
\begin{eqnarray}
	\bar{M}&=& \text{Re}[u(t)e^{i\omega_{0}t}](\sinh^{2} r -|\alpha|^2),\nonumber\\
	\delta M^2 &=& \text{Im}[u(t)e^{i\omega_{0}t}]^{2} [|\alpha \cosh r-\alpha^{\ast} \sinh r e^{i \phi}|^2 +\sinh^{2}r]\nonumber\\
&&+\text{Re}[u(t)e^{i\omega_{0}t}]^{2}[|\alpha|^{2}+\frac{1}{2} \sinh^{2}(2r)]+ \frac{1-|u(t)|^2}{2} N,
\end{eqnarray}
where $u(t)$ satisfies Eq. \eqref{eq:evolution-CF} with the replacement of $\omega_{0}$ by $\omega_{0}+\gamma$. Applying the Markovian solution $u_\text{MA}(t)=e^{-[\kappa+i(\omega_0+\gamma+\Delta)]t}$ with $\kappa=\pi J(\omega_0+\gamma)$ and $\Delta=\mathcal{P}\int_0^\infty{J(\omega)\over\omega_0+\gamma-\omega}d\omega$, we obtain $\min\delta \gamma\simeq({e^{2\kappa t}-1\over2 N t^2})^{1/2}$ when $\beta=(2\sqrt{N})^{-1}$ and $ \varphi=2\phi$. Getting divergent in the long-time limit, its minimum at $t=\kappa^{-1}$ returns the SNL $e\kappa(2N)^{-1/2}$. Thus, the quantum superiority of the scheme in the Markovian noise disappears completely, which is consistent with the result in the Ramsey spectroscopy  \cite{Huelga1997}.

\begin{figure}
\centering
  \includegraphics[width=1\textwidth]{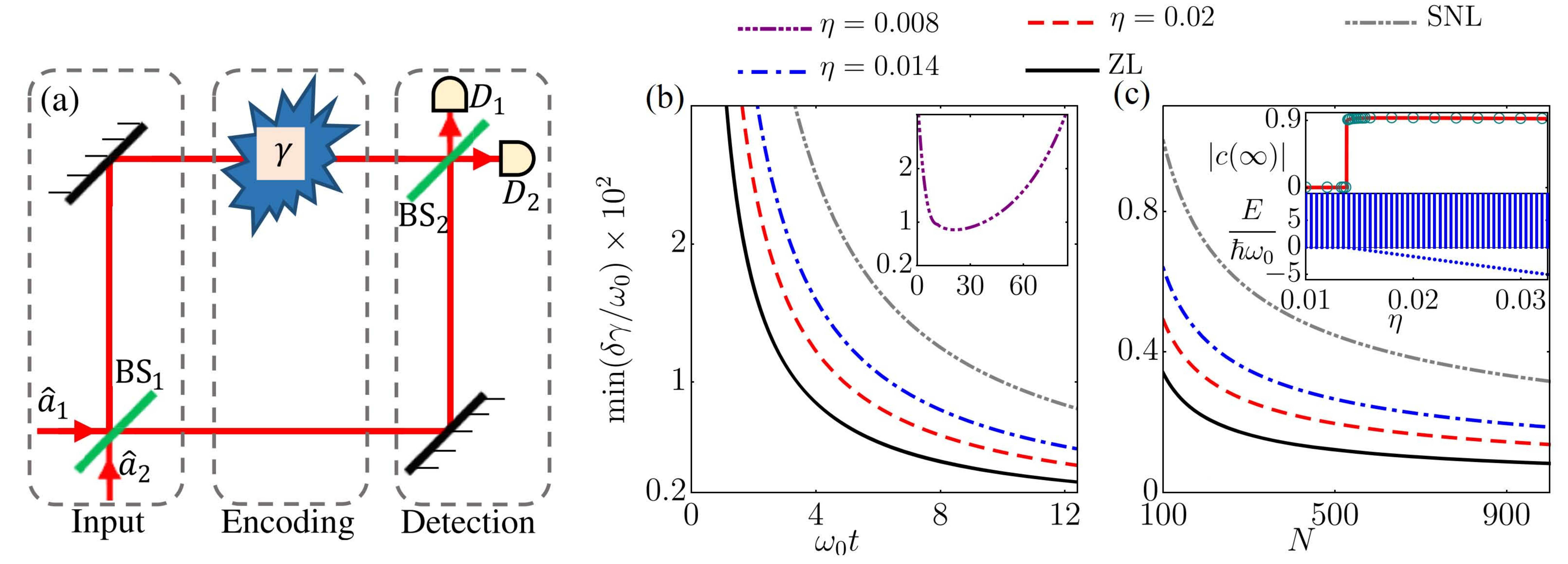}\\
  \caption{(a) Scheme of Mach-Zehnder interference. Metrology precision as a function of time $t$ (b) and total photon number (c) in different $\eta$ of the Ohmic spectral density. The inset of (c) shows steady-state value $Z=|c(\infty)|$ and the energy spectrum. Parameters are $\beta=(2\sqrt{N})^{-1}$, $\omega_{c}=300\omega_{0}$, $\gamma=\pi \omega_{0}$, $t=10\omega_0^{-1}$ for (c) and $N=100$ for (b). Reproduced figures from \cite{Bai2019}. }
  \label{fig:MZI}
\end{figure}

 In the non-Markovian case, focusing on the situation in the presence of the bound state, we can obtain the precision using the long-time solution $Ze^{-iE_bt}$ of Eq. \eqref{eq:eq13}\begin{equation}
\min\delta\gamma|_{\beta=(2\sqrt{N})^{-1}}={(tN^{3/4})^{-1}\over Z}[1+{1-Z^2\over 2Z^2}N^{1\over 2}]^{1\over2},\label{PDLT}
\end{equation}when $t={(2m+1)\pi \over2|\omega_0-\varpi_\text{b}|}$ and $\varphi=2\phi$. It is interesting to see that, different from the previous results \cite{Huelga1997, Escher2011, Matsuzaki2011, Chin2012, Macieszczak2015}, where $\delta\gamma$ gets divergent with time, $\delta\gamma$ decreases with the encoding time monotonically. This is numerically confirmed by the results in Fig. \ref{fig:MZI}(d). Thus, the bound state makes the superiority of the encoding time as a resource in the ideal metrology case retrieved. The relation of the precision with the total photon number in Fig. \ref{fig:MZI}(c) reveals that, with the formation of the bound state, not only the SNL can be surpassed, but also the ideal Zeno limit is asymptotically retrieved with $Z$ approaching unit [see Eq. \eqref{PDLT}].

\subsubsection{Quantum metrology based on Ramsey spectroscopy}
Another widely used metrology scheme is based on the atomic Ramsey spectroscopy \cite{Huelga1997}. To estimate atomic frequency $\omega_0$, we use $n$ two-level atoms prepared in $|\Psi_\text{in}\rangle=(|g\rangle^{\otimes n}+|e\rangle^{\otimes n})/\sqrt{2}$ as the probe. It freely evolves to $|\Psi_{\omega_0}\rangle=(|g\rangle^{\otimes n}+e^{-in\omega_0t}|e\rangle^{\otimes n})/\sqrt{2}$. After performing a CNOT gate with the first atom as the controller and the others as the target to disentangle them and a Hadamard gate on the first atom, the measuring of the excited-state population $\hat{M}=\hat{\sigma}_+\hat{\sigma}_-$ on the first atom results in $\bar{M}=\sin^2{n\omega_0t\over2}$ and $\Delta M=|\sin{n\omega_0t}|/2$. Repeating the experiment in duration $T$, we have $N=T/t$ results. According to the central limit theorem, we have $\delta M=\sqrt{\Delta M\over N}$. Then the propagation of error leads to the best error in estimating $\omega_0$ as
\begin{equation}
\delta \omega_0|_\text{ideal}=(n^2Tt)^{-1/2},\label{HL}
\end{equation}
which is called HL. It has a $n^{1/2}$ time enhancement to the SNL.

Same as the MZI, it is important to estimate the performance of Ramsey interferometry under decoherence. Consider that the atomic free evolution in the parameter encoding step is obscured by a dissipative environment. Thus, the dynamics of each atom is governed by Eq. \eqref{masteqd} with $\hat{o}=\hat{\sigma}_-$. The state after the encoding process reads
\begin{equation}
\rho(t)=\frac{1}{2}\big\{\big{[}|u(t)|^{2}|e\rangle\langle e|+(1-|u(t)|^{2})|g\rangle\langle g|\big{]}^{\otimes n}+|g\rangle\langle g|^{\otimes n}+[u(t)^{n}|e\rangle\langle g|^{\otimes n}+\mathrm{H}.\mathrm{c}.]\big\}.
\end{equation}
Then repeating the same readout process as the ideal case, we can evaluate \cite{Wang2017}
\begin{eqnarray}
\delta \omega_0=\left\{{T[\partial _{\omega_0}\text{Re}(u(t)^n)]^2 \over t[ 1-\text{Re}^2(u(t)^n)]}\right\}^{-1/2}.\label{omegadent}
\end{eqnarray}  Using the Markovian solution $u_\text{MA}(t)=e^{-[\kappa+i(\omega_0+\Delta)]t}$, we obtain that $\delta \omega_0$ tends to be divergent in the long-time limit. After optimizing $t$ in short-time condition, we have the best precision $\min(\delta\omega_0)=(nT/\kappa e)^{-1/2}$, which returns to the SNL. Thus, the quantum superiority of the scheme in the Markovian environment disappears completely \cite{Huelga1997}.

In the non-Markovian dynamics, by substituting the form $Ze^{-iE_bt}$ of large-time $u(t)$ in the presence of the bound state into Eq. (\ref{omegadent}), we have
\begin{equation}
\min(\delta \omega_0)= Z^{-(n+1)}(n^2Tt)^{-1/2},\label{bdsprc}
\end{equation}where the dependence of $E_b$ on $\omega_0$ has been considered via $\partial_{\omega_0}E_b=Z$. Once again, the bound state causes the encoding time as a metrology resource recovered. The precision asymptotically approaches the ideal HL (\ref{HL}) for $n\ll \lfloor -1/\ln Z\rfloor$ when $Z$ reaches unity.

In summary, we find that whether the superiority of the quantum metrology under dissipation exists or not is highly depends on the formation of the bound state and the non-Markovian effect. When the decoherence is Markovian or the bound state is absent, the quantum superiority is destroyed; whereas when the decoherence is non-Markovian and the bound states are formed, the quantum superiority is retrieved. The result suggests a guideline to experimentation to implement the ultrasensitive measurement in the practical noise situation by engineering the formation of the bound state.

\subsection{Quantum plasmonics} \label{sec:decoh-MN}

The hybrid system composed of the metal-dielectric interface and quantum emitters (QEs) has attracted much attention due to its application in modern science, ranging from physics and chemistry to materials science \cite{Kabashin2009,Atwater2010,Giannini2011,Tame2013,Lee2015}. The light is confined in the mode of surface plasmon polariton (SPP), which is a hybrid mode of the electromagnetic field and charge density wave called surface plasmon near the metal-dielectric interface \cite{Chang2006,Pitarke2006,Cacciola2014,AberraGuebrou2012,ToermaeBarnes2014,GonzalezTudela2013}. The fascinating effect of SPP comes from its unique dispersion relation $k_{\text{SPP}}=\frac{\omega}{c}\sqrt{\frac{\varepsilon_\text{m} (\omega)\varepsilon_\text{d}}{\varepsilon_\text{m}(\omega)+\varepsilon_\text{d}}}$, where $\varepsilon_\text{d}$ is the relative permittivities of the dielectric. The metal has a permittivity denoted by a complex Drude model $\varepsilon_\text{m} (\omega )=\varepsilon _{\infty }-\omega _{p}^{2}/[\omega (\omega +i\gamma _{p})]$, where $\omega _{p}$ is the bulk plasma frequency, $\varepsilon _{\infty}$ is the high-frequency limit of $\varepsilon_{\text{m}}(\omega)$, and $\gamma _{p}$ is the Ohmic loss of light in the metal \cite{Barnes2006,EdwardD.Palik1985}. The dispersion relation shows that the SPP has a shorter wavelength than the light. It makes that the SPP can confine light within regions far below the diffraction limit. This effect endows SPP an idea platform to study sub-wavelength optics \cite{Barnes2003,Barnes2006} and strong light-matter interactions \cite{Truegler2008, Wersaell2017, Baranov2018, Chikkaraddy2016, Matsuzaki2017, Kewes2018, Vasa2018}.

\begin{figure}
	\centering
    \includegraphics[width=1\textwidth]{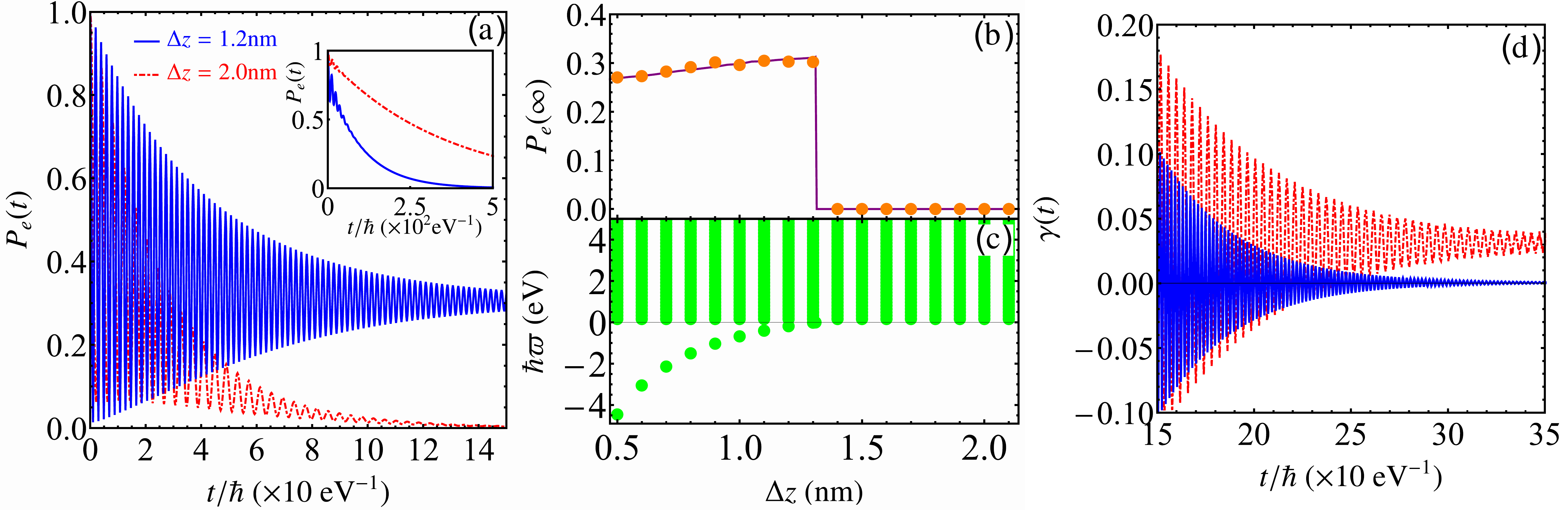}
    \caption{(a) Evolution and (b) long-time values of the excited-state population in different $\Delta z$. The purple line in (b) shows $Z^2$ evaluated by the bound state. (c) Energy spectrum of the whole system in different $\Delta z$. (d) Decay rate of the QE in different $\Delta z$. The parameters are chosen as $\hbar\omega_0=1.2$ eV, $\hbar\gamma_0=10^{-4}$ eV, and $\varepsilon_\text{d}=25$. The inset of (a) shows the evolution when $\hbar\gamma_0=10^{-6}$ eV. Reproduced figures from \cite{Yang2017}.}
    \label{fig:SPP_fig1}
\end{figure}
The quantized QE-SPP interaction is studied in Ref. \cite{Gruner1996} based on the macroscopic quantum electrodynamics. In the vicinity of the nanoparticle-dielectric interface, the spectral density of the SPPs shows a Lorentzian lineshape \cite{Waks2010}. Thus, the SPPs is described by an artificially damping cavity. Based on this assumption, a lot of studies on QE-metal nanoparticle hybrid system have been given \cite{Shah2013, Delga2014, Ge2015, Li2016, Peng2017}. However, the Lorentzian spectral density approximation only correctly captures the short-time dynamics of the QEs. On the other hand, due to absorption of the metal to the SPPs, QE coupled to the SPPs tends to its ground state in a long-time limit. It is the main obstacle in the practical application of a hybrid QE-SSP system. Therefore, how to stabilize the coherence of QE is an urgent problem. Recent studies have shown that the formation of the QE-SPP bound states could supply an efficient way to suppress the decoherence of the QE and the damping of the SPPs in the metal \cite{Yang2017,Yang2019}.

Consider a QE with frequency $\omega_0$ placed at a distance $\Delta z$ above the silver interface. The Hamiltonian follows Eq. \eqref{eq:Hamiltonian} with the $\bf{k}$-th SPP mode-emitter interaction strength $g_{\bf{k}}=\frac{ic^{-2}\omega_{\bf{k}}^2}{\sqrt{\pi \varepsilon_0/\hbar}}\int d^3\mathbf{r}' \sqrt{\text{Im}[\varepsilon_m(\omega_{\bf{k}})]}\bm{\mu}\cdot \mathbf{G}(\mathbf{r},\mathbf{r^\prime},\omega_{\bf{k}}) \cdot \bf{e}_{\bf{k}}$. Here $\varepsilon_0$ being the vacuum permittivity, $c$ the speed of light, $\bf{G}(\bf{r},\bf{r}^{\prime},\omega)$ is the Green's function satisfying the Helmholtz equation $[\nabla\times\nabla\times -\omega^2\varepsilon_m(\omega)/c^2]\mathbf{G}(\mathbf{r},\mathbf{r}',\omega)=\delta(\mathbf{r}-\mathbf{r}')$, $\bf{r}$ is the position of the emitter with electric dipole $\bm{\mu}$, and $\bf{e}_{\bf{k}}$ is the polarization direction of the SPP mode. Taking the dipole $\bm{\mu}$ along the $z$ direction, the spectral density of the SPPs reads \cite{Yang2017}
\begin{equation}
J(\omega)={3\gamma_0\sqrt{\varepsilon_\text{d}}\omega^3\over 4\pi \omega_0^3}\text{Re}[\int_0^\infty ds{s^3(1-r_\text{p}e^{2ik_{z_\text{d}}\Delta z})\over \sqrt{1-s^2}}],\label{fspec}
\end{equation}
where $s=k_{\rho}/k_{d}$, $r_{\text{p}}=\frac{\varepsilon _{\text{d}}k_{z_{\text{m}}}-\varepsilon _{\text{m}}(\omega )k_{z_{\text{d}}}}{\varepsilon _{\text{d}}k_{z_{\text{m}}}+\varepsilon _{\text{m}}(\omega )k_{z_{\text{d}}}}$ with $k_{z_{\text{m}}}$ and $k_{z_{\text{d}}}$ being the axial components of the wave vector in the metal and the dielectric, respectively, and $\gamma_0=\omega_0^3 \mu/(3\pi\hbar\varepsilon_0 c^3)$ is the vacuum spontaneous emission rate. The evolution of the QE obeys Eq. \eqref{masteqd}. Figure \ref{fig:SPP_fig1} shows the evolution of the excited-state population $P_e(t)=\text{Tr}[\hat{\sigma}_+\hat{\sigma}_-\rho(t)]$ of the QE in different $\Delta z$. When the intrinsic decay rate $\gamma_0$ is small, $P_e(t)$ monotonically decays to zero, especially for large $\Delta z$ [see the inset of Fig. \ref{fig:SPP_fig1}(a)]. When $\gamma_0$ is large, the dynamical oscillation becomes significant. It represents a rapid excitation exchange and thus manifests the strong QE-SPP interactions. A remarkable difference between the two results is that although $P_e(t)$ approaches zero when $\Delta z=2.0$nm, while it tends to a stable nonzero value when $\Delta z=1.2$nm. It is in sharp contrast to the previous result and represents a dissipation suppression of the QE. Figure \ref{fig:SPP_fig1}(b) shows the steady-state population $P_e(\infty)$ in different $\Delta z$. It reveals that $P_e(\infty)$ matches well with $Z^2$ obtained by the bound-state analysis. Figure \ref{fig:SPP_fig1}(c) shows clearly that the regions where the QE population is preserved correspond exactly to the one where a bound state is formed in the environmental bandgap. Figure \ref{fig:SPP_fig1}(d) is the decay rate $\gamma(t)$ of the QE. The fast oscillations manifest the rapid energy exchange between the QE and the SPPs. It is interesting to see that $\gamma(t)$ asymptotically approaches zero with the formation of the bound state. Such behavior implies that the plasmonic nanostructure in the presence of the bound state can be an ideal platform for the coherent manipulation of QE dynamics, even it is inherently dissipative. This is helpful for utilizing plasmonic nanostructures in quantum devices, such as entanglement generators \cite{Gonzalez-Tudela2011}.

The formation of the bound state in the QE-SPPs system also indicates that the SPPs can be used to realize entanglement preservation \cite{Yang2019}.

\section{Quantum control in periodically driven systems}\label{sec:Floquet}

Coherent control via periodic driving dubbed as Floquet engineering has become a versatile tool in artificially synthesizing exotic states of matter in systems of ultracold atoms \cite{Eckardt2017,Meinert2016}, photonics \cite{Rechtsman2013,Cheng2019}, superconductor qubits \cite{Roushan2017}, and graphene \cite{McIver2020}. Different from the static systems, the systems under periodic driving has no well-defined energy spectrum because its energy is not conserved. Floquet theory \cite{Shirley1965, Sambe1973} supplies us with a powerful approach to map a nonequilibrium system under periodic driving to a static one. According to Floquet theory, the periodic system with $\hat{H}(t+T)=\hat{H}(t)$ has a complete set of basis $|\phi_{\alpha }(t)\rangle $ determined by
 \begin{equation}
\lbrack\hat{H}(t)-i\partial_t]|\phi_{\alpha}(t)\rangle =\epsilon_{\alpha }|\phi_{\alpha }(t)\rangle\label{flqe}
\end{equation}
such that the evolution of any state can be expanded as $|\Psi(t)\rangle=\sum_\alpha C_\alpha e^{-i \epsilon_\alpha t} |\phi_\alpha(t) \rangle$ with $C_\alpha=\langle u_\alpha(0)|\Psi(0)\rangle$. The time-independence of $C_\alpha$ implies that $\epsilon _{\alpha}$ and $|\phi_{\alpha}(t)\rangle$ play the same roles in a periodic system as eigenenergies and stationary states do in a static system. They are thus called quasienergies and quasi-stationary states, respectively. Carrying all the quasi-stationary-state characters, the quasienergy spectrum formed by all $\epsilon _{\alpha}$ is a key to study the periodic system. Note that $\epsilon _{\alpha}$ is periodic with period ${2 \pi/ T}$ because $e^{i l\omega t} |u_\alpha(t)\rangle$ with $\omega={2\pi/T}$ is also the eigenstate of Eq. (\ref{flqe}) with eigenvalue $\epsilon_\alpha+l\omega$. It can be proven that the Floquet eigenequation \eqref{flqe} is equivalent to \begin{equation}
\hat{U}_T|\phi_\alpha(0)\rangle=e^{-i\epsilon_\alpha T}|\phi_\alpha(0)\rangle \label{oprdegf}
\end{equation} with $\hat{U}_T$ the one-period evolution operator. Thus, the Floquet equation defines an effective static system $\hat{H}_\text{eff}=\frac{i}{T}\ln \hat{U}_T$ whose eigenvalues are the quasienergies. Then one can engineer the quasienergy spectrum by changing the parameters of the external periodic driving and artificially synthesize many exotic states of matter absent in the static systems. Its merits reside in that the controllability of the systems is dramatically enhanced because time as a new control dimension is introduced to the systems by the periodic driving. This greatly lifts the experimental difficulty in realizing different types of quantum states of matter via changing the intrinsic parameters of the systems, which is hard to be manipulated once the material of the system is fabricated.

In the above section, we reviewed the decoherence control via engineering the bound state in the energy spectrum. A practical difficulty is that one generally cannot change the energy spectrum once the system is explicitly given. In this section, we will review that a parallel decoherence control can be realized by periodic driving as long as a Floquet bound state (FBS) is formed in the quasienergy spectrum. Actually, such engineering to the FBS in the quasienergy spectrum can be generalized to realize artificial topological phases. We will also review its recent progress.

\subsection{Floquet control of quantum dissipation}\label{sec:Floquet-BS}

\begin{figure}
	\centering
    \includegraphics[width=1\textwidth]{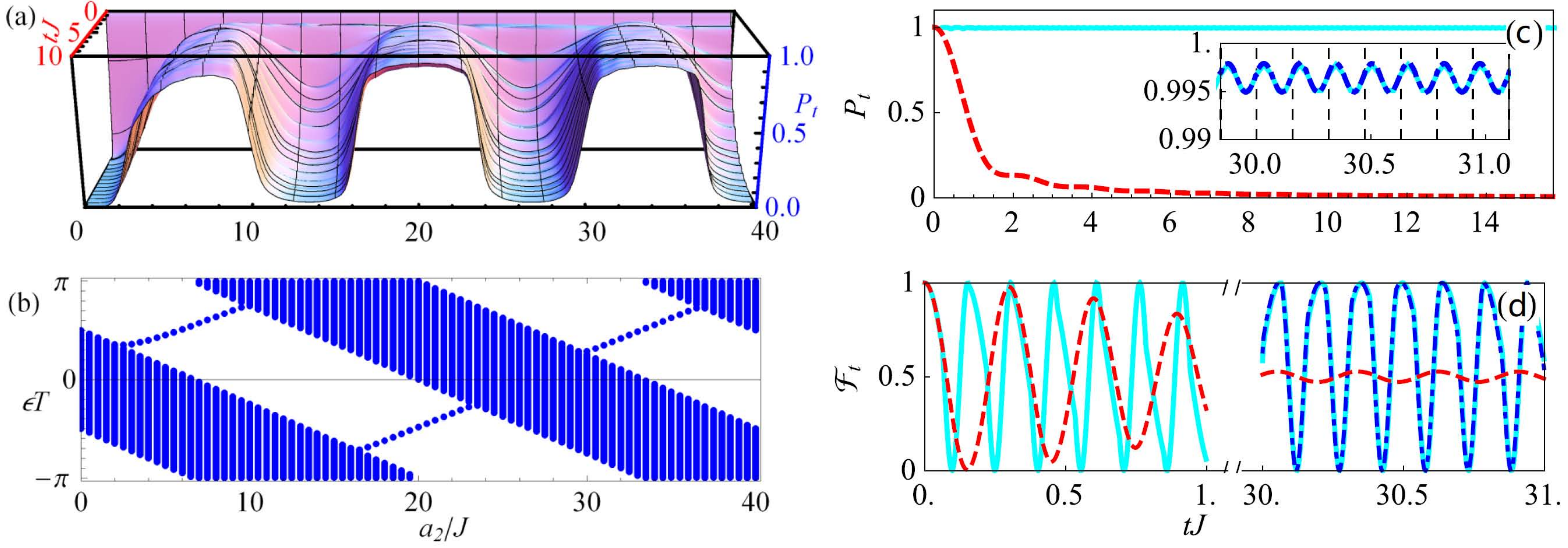}
    \caption{(a) Evolution of $P_t$ and (b) quasienergy spectrum of the whole system in different driving amplitude $a_2$. The parameters $T=0.25\pi J^{-1}$, $a_1=0$, $\tau=0.1\pi J^{-1}$, $g=1.0J$, $\lambda=20.0J$, and $L=800$ are used. Evolution of $P_t$ for the initial state $|\phi\rangle=|\uparrow_0\rangle$ in (c) and $\mathcal{F}_t$ for the initial state $|\phi\rangle=(|\uparrow_0\rangle+|\downarrow_0\rangle)/\sqrt{2}$ in (d) when $a_2=36.0J$ with the FBS (cyan solid line) and $a_2=1.5J$ without the FBS (red dashed line). The blue dotdashed lines show the results obtained via analytically evaluating the contribution of the FBS to the asymptotic state. The parameters in (c, d) are $T=0.05\pi J^{-1}$ and $\tau=0.02\pi J^{-1}$. Reproduced figures from \cite{Chen2015}.} \label{fig:FBS}
\end{figure}

The idea of controlling quantum dissipation via periodic driving is based on the correspondence between the quasienergy spectrum in a temporally periodic system and the energy spectrum in the static system \cite{Chen2015}. Compared with the static case, the periodic driving provides a much flexible tool in manipulating the formation of the bound state.

Consider a periodically driven spin-1/2 system interacting with an XX-coupled spin chain environment. The Hamiltonian reads $\hat{H}(t)=\hat{H}_{S}(t) + \hat{H}_{I}+\hat{H}_{E}$ with
\begin{eqnarray}
	\hat{H}_{S}(t)&=&\frac{1}{2}[\lambda+A(t)]\hat{\sigma}_0^z,~~ \hat{H}_{I}= \frac{g}{2}(\hat{\sigma}_0^x\hat{\sigma}_1^x+\hat{\sigma}_0^y\hat{\sigma}_1^y),\nonumber\\
	\hat{H}_{E}&=&\frac{\lambda}{2}\sum_{j=1}^L \hat{\sigma}_j^z+\frac{J}{2}\sum_{j=1}^{L-1}(\hat{\sigma}_j^x \hat{\sigma}_{j+1}^x+\hat{\sigma}_j^y \hat{\sigma}_{j+1}^y).
\end{eqnarray}
Here $\hat{\sigma}^{\alpha}_{j}$ ($\alpha=x,y,z$) being the Pauli matrix of the $j$-th spin, $\lambda$ is the Zeeman interaction induced by the longitudinal magnetic field, $J$ and $g$ are the coupling strengths between the nearest-neighbor spins in the spin chain and the coupling between system spin and the first spin of the chain, respectively, and $L$ is the total number of spins in XX-chain. Here the periodic driving field reads
\begin{equation}
	A(t)=\begin{cases}
		a_1, & n T<t \le nT+\tau \\
		a_2, & nT+\tau <t \le (n+1)T
 	\end{cases},
\end{equation}
where $T$ is the driving period, $\tau$ is the duration of the first piece, and $a_{1,2}$ are the driving amplitudes. We consider the initial state as $|\Psi(0)\rangle=|\uparrow_0\rangle\otimes|\{\downarrow_j\}\rangle$. Because the total excitation number $\hat{\mathcal{N}}=\sum_{j=0}^L\hat{\sigma}_j^+ \hat{\sigma}_j^-$ is conserved, the time-evolved state can be expanded as $|\Psi(t)\rangle=\sum_{j=0}^Le^{i \frac{L\lambda t}{2}}c_j(t)\hat{\sigma}_j^+|\{\downarrow_j\}\rangle$. Here the excited-state population $P_t\equiv |c_0(t)|^2$ of the system spin is used to characterize the decoherence effect.

Figure \ref{fig:FBS}(a) shows the evolution of $P_t$ in different driving amplitude $a_2$. When the driving is switched off, i.e., $a_2=0$, $P_t$ decays monotonically to zero, which means a complete dissipation exerted by the spin chain to the system spin. When the driving is switched on, it is interesting to see that $P_t$ is stabilized repeatedly with the increase of $a_2$. It is remarkable to see that the regimes where the dissipation is inhibited match well with the ones where the FBS is formed in the quasienergy spectrum [see Fig. \ref{fig:FBS}(b)]. To understand the behavior, we, according to Floquet theory, rewrite $|\Psi(t)\rangle$ as
\begin{eqnarray}|\Psi(t)\rangle&=&e^{i{\frac{ L \lambda t}{2}}}[xe^{-i\epsilon_\text{FBS}t}|\phi_\text{FBS}(t)\rangle+\sum_{\alpha\in\text{Band}}y_\alpha e^{-i\epsilon_\alpha t}|\phi_\alpha(t)\rangle],\end{eqnarray} where $x=\langle \phi_{\text{FBS}}(0)|\Psi (0)\rangle $ and $y_{\alpha }=\langle \phi_{\alpha }(0)|\Psi (0)\rangle $. One can see that $P_t$ evolves asymptotically to $P_\infty\equiv x^2|\langle\Psi(0)|u_\text{FBS}(t)\rangle|^2$ with all the components in the quasi-energy band vanishing due to the out-of-phase interference contributed by the continuous phases, as confirmed in Fig. \ref{fig:FBS}(c). In the absence of the FBS, $P_t$ decays to zero finally. It means that the presence of the FBS would cause $P_t$ to survive in the only component of the FBS and thus synchronize with the driving field \cite{Russomanno2012}.
For a general initial state $|\phi(0)\rangle=\frac{1}{\sqrt{2}}(|\uparrow_0\rangle+|\downarrow_0\rangle)$, we use the initial-state fidelity $\mathcal{F}_{t}=\langle \phi(t)|\text{Tr}_{E}[|\Psi(t)\rangle\langle \Psi(t)|]|\phi(0)\rangle$ to reflect the damping of the quantum coherence. The dynamics are shown in Fig. \ref{fig:FBS}(d). In the absence of the FBS, $\mathcal{F}_{t}$ decays to a trivial value $1/2$, which indicates that the system $\rho_s=\text{Tr}_{E}[|\Psi(t)\rangle\langle \Psi(t)|]=\frac{\mathbb{I}}{2}$ has no coherence, while $\mathcal{F}_{t}$ shows persistent oscillation once the FBS is formed. The long-time dynamics in Fig. \ref{fig:FBS}(d) verifies again the coincidence of the FBS analysis (blue dashed line) with the exact numerical results (cyan line).

As a stationary state, the bound state in both the static system and the periodically driven system makes the system survive in the only component of the bound state in the long-time limit. Thus, the coherence contained in the bound state is preserved forever.  Different from the bound state in a static system, the FBS formed in the quasienergy spectrum itself is temporally periodic. Such periodic driving induced dissipation suppression can be seen as a time-dependent analog to the decoherence suppression induced by the structured environment in spatially periodic photonic crystal setting \cite{PhysRevLett.107.193903}, where the spatial periodicity causes the environment to present a bandgap structure and thus the decoherence of the system is suppressed when the system frequency resides in the bandgap region. However, the temporally periodic driving relaxes greatly the experimental difficulties of the static system in fabricating specific spatial periodicity. Such an FBS mechanism is an exact non-Markovian dynamical control to suppress quantum dissipation by using periodic driving. An optical simulation of this mechanism has been proposed in Ref. \cite{Ma2018}.

\subsection{Floquet control of dissipative quantum battery}\label{sec:Floquet-QB}

A quantum battery (QB) is a device that stores and converts energy at the quantum level \cite{Alicki2013}. It holds the promises of higher energy-storing density in large-scale integration and faster-charging power than its classical counterpart. A minimal model of QB is described by two coupled two-level systems with one acting as the charger and another as the QB \cite{Ferraro2018,Andolina2019}. The Hamiltonian reads
\begin{equation}
\hat{H}_{0}(t)=\hbar\sum_{l=\rm{b}, \rm{c}} \omega_{l} \hat{\sigma}^{\dagger}_{l}\hat{\sigma}_{l} + \hbar\kappa f(t) (\hat{\sigma}^{\dagger}_\text{b} \hat{\sigma}_\text{c}+\text{H.c.}),
\label{eq:Qbattery}
\end{equation}
where $\hat{\sigma}_l=|g\rangle_l\langle e|$, with $|g\rangle$ and $|e\rangle$ being the ground and excited states of the two-level systems, are the transition operators of the QB and charger with frequency $\omega_l$, and $\kappa$ is their coupling strength. The time-dependent $f(t)$ is a control parameter for the charging, storing, and discharging process that takes
\begin{equation}
f(t)=\left\{\begin{array}{ll}
1, & nT<t \le \tau_c+nT\\
0, & nT+\tau_c<t\le nT+(\tau_s+\tau_c)\\
1, & nT+(\tau_s+\tau_c)<t\le (n+1)T\\
\end{array} \right.
 n\in \mathbb{N}.
\label{eq:Control-QB}
\end{equation}
Here $\tau_c$, $\tau_s$, and $\tau_d$ are the time durations of charging, storing, and discharging, respectively. The initial state of the QB-charger setup is set to $|\psi(0)\rangle=|g_{\rm{b}}, e_{\rm{c}}\rangle$. The average energy $\mathcal{E}(t)=\omega_{\rm{b}}\langle\psi(t)| \hat{\sigma}^{\dagger}_{\rm{b}}\hat{\sigma}_{\rm{b}}|\psi(t)\rangle$ of the QB is used to characterize its power, where $|\psi(t)\rangle \equiv \mathcal{T} e^{{-i\over\hbar}\int_0^t \hat{H}_0(\tau)d\tau}|\psi(0)\rangle$ is the evolved state of the QB-charger setup. For the resonant case $\omega_{\rm{b}}=\omega_{\rm{c}}$, with the optimal control time, i.e. $\tau_c=\tau_s=\tau_d=\frac{2\pi}{\kappa}$ \cite{Bai2020}, the QB-charger setup shows the perfect cyclic evolution: $\mathcal{E}(t)$ reaches its maximal value at the end of each charging step, and empties its total energy at the end of each discharging step. However, in the realistic condition, an invertible power loss is present due to the interactions of the QB and the charger with their respective dissipative environment. It was found that the dissipation causes that energy of QB drop to zero after finite charing-storing-discharging cycles, which is called the aging of the QB \cite{Pirmoradian2019,Kamin2020}. The temporally periodical nature of the cyclic process of the QB inspires us that the detrimental effect of the dissipation on the QB-charger setup can be inhibited by engineering the formation of the FBSs. Then, we review how to reactive the QB under the influences of the dissipative environments by manipulating the FBSs \cite{Bai2020}.

\begin{figure}
	\centering
    \includegraphics[width=0.4\textwidth]{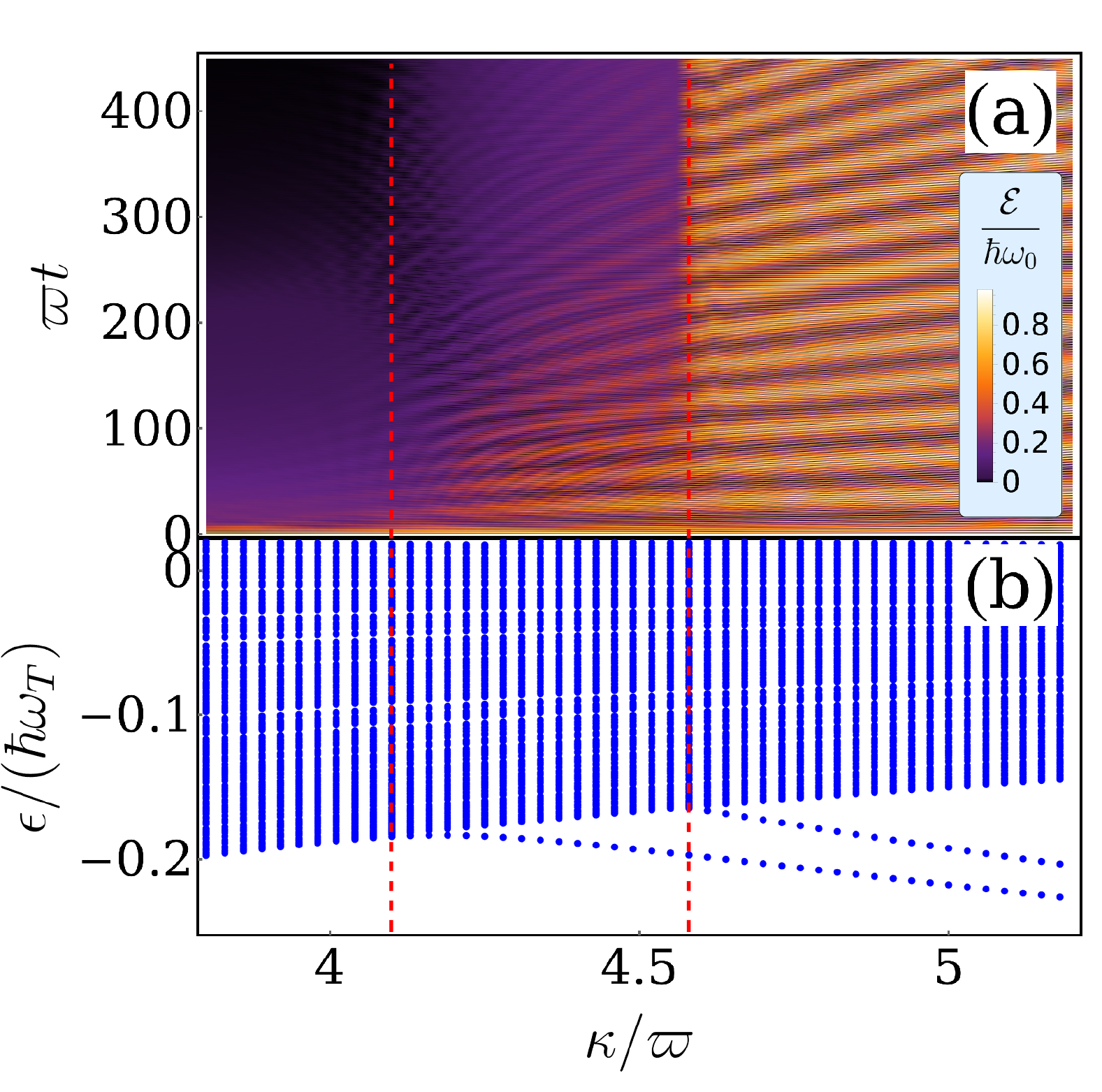}~\includegraphics[width=0.6\textwidth]{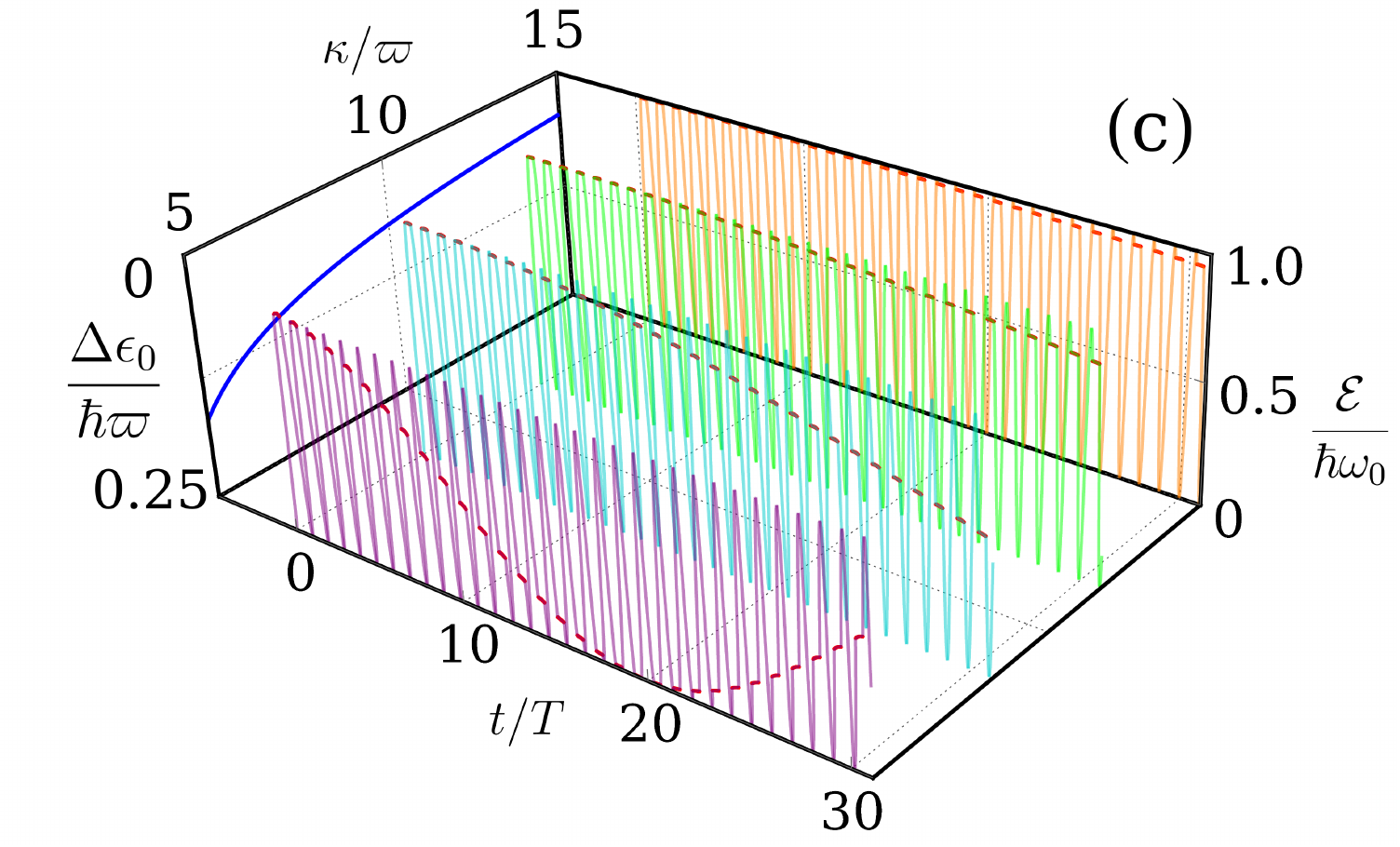}
      \caption{(a) Evolution of the energy of the QB and (b) quasienergy spectrum of the total system in different coupling strength $\kappa$. (c) Evolution of $\mathcal{E}(t)$ in long-time limit and the quasienergy difference $\Delta \epsilon_0$ of two FBSs in different $\kappa$. The energy in storage time duration is highlighted by red segments in (c). Other parameters are $N=30$, $\omega_b=\omega_c=2\varpi$, $g=0.5\varpi$,  $q=0.5\varpi$ and $\tau_c=\tau_s=\tau_d=\pi/(2\kappa)$. Reproduced figures from \cite{Bai2020}. } \label{fig:QB}
\end{figure}

Consider that the QB and the charger are coupled to two independent environments consisting of an $N\times N$ nearest-neighbor coupled bosonic mode on the square lattice with the mode frequency $\varpi$ and hopping rate $q$ \cite{GonzalezTudela2017}. Their couplings follow Eq. \eqref{eq:Hamiltonian} with $\hat{o}\equiv\hat{\sigma}_{\rm{b},(\rm{c})}$, $\omega_{\mathbf{k}}\equiv \varpi-2q(\cos k_x+\cos k_y)$, and the system-environment coupling strength being $g_{\mathbf{k}}=\frac{g}{N}$. Figure \ref{fig:QB}(a) and (b) show the evolution of $\mathcal{E}(t)$ in different coupling strength $\kappa$ and the corresponding quasienergy spectrum. We can see that the
regions where different numbers of FBSs are formed match well with the ones where $\mathcal{E}(t)$ shows different behaviors, i.e. complete decay, energy trapping,
and persistent oscillation. When no FBS is formed, $\mathcal{E}(t)$ asymptotically decays to zero. However, when one or two FBSs are formed, $\mathcal{E}(t)$ approaches finite value or shows a persistent oscillation in the long-time limit. According to the Floquet theory, such anomalous dynamical behaviors can be understood by decomposing the evolution state of the total system as $|\Psi(t)\rangle=\sum_{j=1}^Mc_je^{{-i\over\hbar} \epsilon_{0j}t}|\phi_{0j}(t)\rangle+\sum_{\beta\in\text{CB}}d_{\beta}e^{{-i\over\hbar}\epsilon_{\beta}t}|\phi_{\beta}(t)\rangle $, where $\epsilon$ and $|\phi(t)\rangle$ satisfy Eq. \eqref{flqe}, $M$ is the number the FBSs, $c_j\equiv\langle \phi_{0j}(0)|\Psi(0)\rangle$ and $d_{\beta}\equiv\langle \phi_\beta(0)|\Psi(0)\rangle$. In the long-time limit, the contribution from the continuous energy band to $\mathcal{E}(t)$ approaches zero due to the out-of-phase interference. Then, $\mathcal{E}(\infty)$ reads
\begin{eqnarray}
{\mathcal{E}(\infty)\over\hbar\omega_0}=\sum_{jj'=1}^Mc_jc_{j'}^*e^{{-i\over\hbar}(\epsilon_{0j}-\epsilon_{0j'})t}\langle \phi_{0j'}(t)|\hat{\sigma}_\text{b}^\dag\hat{\sigma}_\text{b}|\phi_{0j}(t)\rangle,
\end{eqnarray}
which indicates that the steady behavior of $\mathcal{E}(\infty)$ strongly dependent on the number of FBSs, as shown in Fig. \ref{fig:QB}(a) and (b). In particular, when $M=2$, $\mathcal{E}(\infty)$ shows the persistent oscillation with multiple frequencies jointly determined by the driven frequency $\omega_T$ and the quasienergy difference of two FBSs $\Delta \epsilon_0\equiv |\epsilon_{01}-\epsilon_{01}|$.  With the bound-state mechanism at hands, it is shown that the QB influenced by the environment can be reactivated by controlling the energy of two FBSs. As shown in Fig. \ref{fig:QB}(c), with the increasing of $\kappa$, the quasienergy difference asymptotically tends to zero, which reduces the undesired energy oscillation contributed from the interference between two FBSs and thus the QB returns to its near-ideal cyclic stage.

The combination between the decoherence analysis of QB and the Floquet theory supplies an efficient way to overcome the aging problem of the QB via the versatile Floquet engineering. It opens an avenue to build an aging-free QB in practice.


\subsection{Floquet control of topological phases}\label{sec:Floquet-topological}
The distinguished role of the periodic driving in engineering bound states in the quasienergy spectrum can also be used to artificially synthesize exotic topological phases \cite{Rudner2013,Kundu2014,Lababidi2014,Tong2013,Xiong2016,Liu2019,Wu2020}. Parallel to the topological phases in the static system, the topological phases in periodically driven systems are called Floquet topological phases. Floquet topological phases defined in the quasienergy spectrum are characterized by a full insulating gap in the bulk and gapless bound states at the edges or the surfaces \cite{Kitagawa2010}. The versatility of periodic driving schemes enables us not only to realize topological phases not accessible for the static system in the same setting, but also to explore novel topological phases totally absent in the static system.

Consider a two-band topological system with its parameters periodically driven by a pairwise function with the Hamiltonians read $\mathcal{H}_1$ and $\mathcal{H}_2$ in the respective time durations $T_1$ and $T_2$. The coefficient matrices of the Hamiltonians in the operator basis of the momentum space take the form $\mathcal{H}_j(\mathbf{k})=\mathbf{h}_j(\mathbf{k})\cdot\pmb{\sigma}$ ($j=1,2$) with $\pmb{\sigma}$ being the Pauli matrices. The one-periodic evolution operator reads ${U}_T=\varepsilon I_{2\times 2}+i\mathbf{r}\cdot \pmb{\sigma}$ which yield the effective Hamiltonian $\mathcal{H}_\text{eff}(\mathbf{k}):=\frac{i}{T}\ln {U}_T=\mathbf{h}_\text{eff}(\mathbf{k})\cdot \pmb{\sigma}$ with $\mathbf{h}_\text{eff}(\mathbf{k})=-\arccos (\varepsilon)\underline{\mathbf{r}}/T$ and \cite{Xiong2016}
\begin{eqnarray}
\varepsilon&=&\cos|T_1 \mathbf{h}_1(\mathbf{k})|\cos|T_2 \mathbf{h}_2(\mathbf{k})|-\underline{\mathbf{h}}_1\cdot\underline{\mathbf{h}}_2\sin|T_1\mathbf{h}_1(\mathbf{k})|\sin|T_2\mathbf{h}_2(\mathbf{k})|,\nonumber\\
\mathbf{r}&=&\underline{\mathbf{h}}_1\times\underline{\mathbf{h}}_2\sin|T_1\mathbf{h}_1(\mathbf{k})|\sin|T_2\mathbf{h}_2(\mathbf{k})|-\underline{\mathbf{h}}_2\cos|T_1\mathbf{h}_1(\mathbf{k})|\sin|T_2\mathbf{h}_2(\mathbf{k})|\nonumber\\ &&-\underline{\mathbf{h}}_1\cos|T_2\mathbf{h}_2(\mathbf{k})|\sin|T_1\mathbf{h}_1(\mathbf{k})|.\label{var}
\end{eqnarray}
Here $T=T_1+T_2$ and $\underline{\mathbf{v}}\equiv\mathbf{v}/|\mathbf{v}|$ is the unit vector of $\mathbf{v}$. The topological phase transition is associated with the closing and reopening of the quasi-energy bands. One can find from Eqs. \eqref{var} that the band gap closing occurs at $\varepsilon=\pm1$, which is obtained when
\begin{eqnarray}
&&T_j|\mathbf{h}_j(\mathbf{k})|=n_j\pi,~n_j\in \mathbb{Z}, \label{gen}\\
&&\text{or}~
\begin{cases}
\underline{\mathbf{h}}_1\cdot\underline{\mathbf{h}}_2=\pm1\\
T_1|\mathbf{h}_1(\mathbf{k})|\pm T_2|\mathbf{h}_2(\mathbf{k})|=n\pi,~n\in\mathbb{Z}
\end{cases}\label{hh1}
\end{eqnarray}
at the quasienergy zero (or $\pi/T$) if $n$ is even (or odd). They give the sufficient condition for judging the occurrence of the Floquet topological phase transition.

\subsubsection{Floquet Majorana modes}

\begin{figure}
	\centering
    \includegraphics[width=0.6\textwidth]{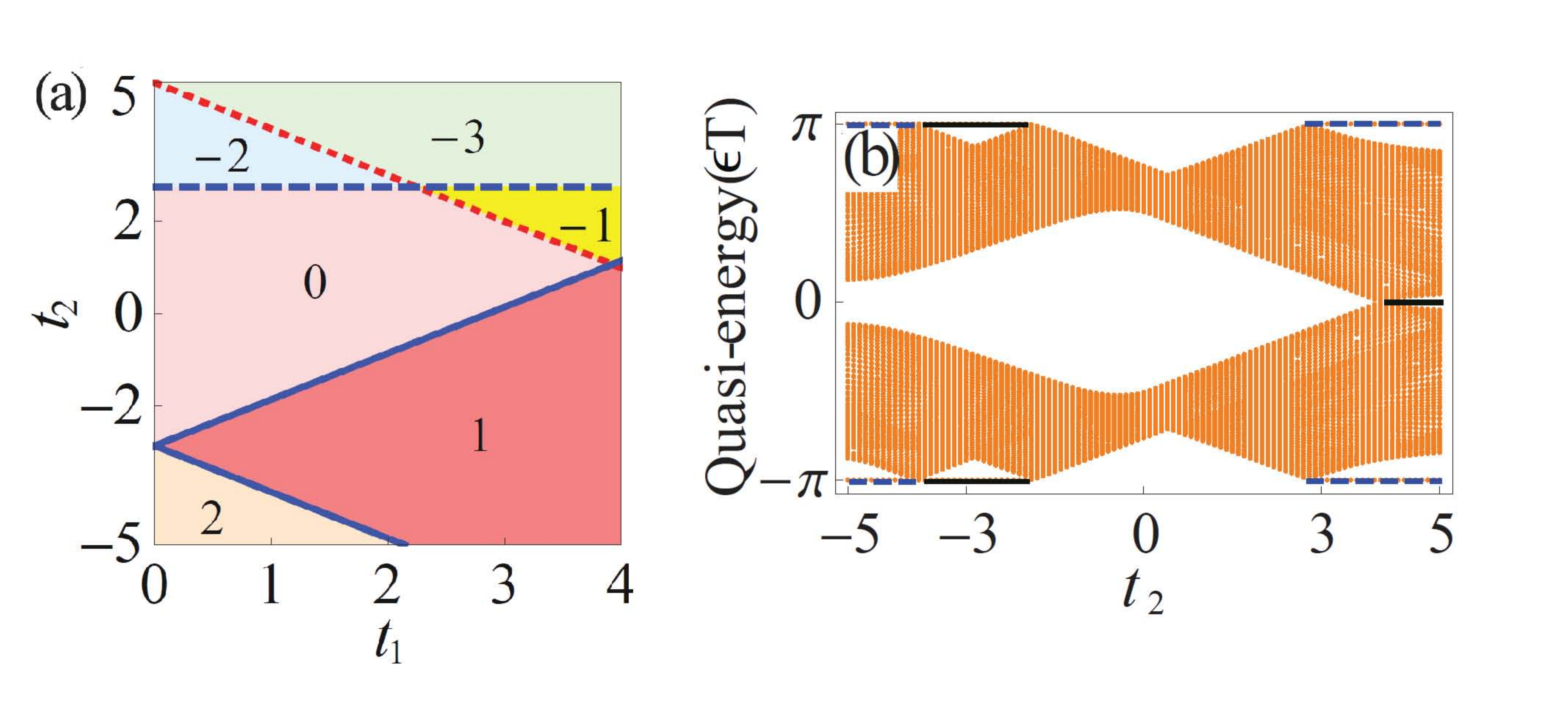}
      \caption{(a) Phase diagram characterized by the winding number when $\mu = -10|\Delta_1|$ and $|\Delta_2|= 2.5|\Delta_1|$. (b) Quasienergy spectrum for $t_1=|\Delta_1|$. The dashed blue line and solid black line stand for two degenerate pairs and a single pair of MMs, respectively. Reproduced figures from \cite{Tong2013}. } \label{fig:fmm}
\end{figure}

Realizing Majorana modes (MMs) in condensed-matter systems is of vast experimental and theoretical interests due to their potential applications in quantum computation \cite{Fu2008,Sau2010,Sato2009,Gong2012,Zhu2011}. The zero-bias conductance peaks are regarded as a signature of MMs in spin-orbit coupled semiconductor nanowires \cite{Williams2012}. However, they can also be induced by other reasons, e.g. the strong disorder in the nanowire \cite{Liu2012,Kells2012} and smooth confinement potential at the wire end \cite{Kells2012}. Therefore, more ways to double confirm the formation of MMs in these systems are desired. Studies show that periodic driving can create a widely tunable number of Floquet MMs, which, on the one hand, could dramatically enhance the experimental signal strength, on the other hand, could supply a novel way to identify whether the signal originates from MFs or others by observing its response to the tuning of the driving coefficients \cite{Tong2013}.

Consider the Kitaev model Hamiltonian for a 1D spinless $p$-wave superconductor
\begin{eqnarray}
H&=&-\mu\sum_{l=1}^{N}c_{l}^{\dag}c_{l}-\sum_{l=1}^{N-1}(t_{1}c_{l}^{\dag}c_{l+1}+\Delta_{1}c_{l}^{\dag}c_{l+1}^{\dag}+\text{h.c.})-\sum_{l=1}^{N-2}(t_{2}c_{l}^{\dag}c_{l+2}\nonumber\\
&&+\Delta_{2}c_{l}^{\dag}c_{l+2}^{\dag}+\text{h.c.}),\label{Hamil}
\end{eqnarray}
where $\mu$ is the chemical potential, $t_{a}$ and
$\Delta_{a}=\left\vert\Delta_{a}\right\vert e^{i\phi_a}$ with $a=1$
($a=2$) describe the nearest- (next-nearest) -neighbor hopping
amplitude and pairing potential, respectively, and $\phi_{a}$ is the
associated superconducting phases. Its topological characters are dependent on the relative phase $\phi=\phi_1-\phi_2$ \cite{Ryu2010}. For $\phi=0$ and $\pi$, $H$ has
time-reversal and particle-hole symmetries and its topology is characterized by the winding number, where two pairs of Majorana modes can be generated at most. For other values of $\phi$, $H$ has no time-reversal and particle-hole symmetries and its topology is characterized by a $Z_2$ topological invariant, where one pair of Majorana modes is achievable at most.

The periodic driving protocol is designed as follows: in the first half period, $H_1=H(\phi_1,\phi_2)$ with both $\phi_1$ and $\phi_2$ fixed; whereas in the second half period, we swap $\phi_1$ and $\phi_2$ so that $H_2=H(\phi_2,\phi_1)$. It has been proved that the distinguished action of the periodic driving is that it successfully recovers the time-reversal symmetry and simulates the long-range hopping of the electrons. Thanks to these abilities, an arbitrarily number of MMs is created. Figure \ref{fig:fmm}(a) shows the phase diagram for given driving parameters with the corresponding quasienergy spectrum shown in Fig. \ref{fig:fmm}(b). Compared to the static case which supports at most two pairs of MMs, Floquet engineering provides a feasible method to realize tunable topological edge states. It was reported that more MMs can be induced by increasing the driving period $T$, where longer-range hopping is simulated by the periodic driving \cite{Tong2013}.

\subsubsection{Floquet Chern insulators}
The large-Chern-number topological phases have attracted considerable interests \cite{Wang2013, Skirlo2014, Fang2014, Moeller2015, Scaffidi2015, Roentynen2015, Skirlo2015}. Providing more edge channels, these phases are expected to improve certain device performance by lowering the contact resistance \cite{Wang2013, Fang2014}. They can also support multi-mode one-way waveguides, a feature relevant to realizing reflectionless waveguides splitters, combiners, and one-way photonic circuits in photonic crystals \cite{Skirlo2014, Skirlo2015}. However, a band structure with large Chern numbers is normally hard to be generated in natural material. Special long-range interactions and multi-layered structure of the materials are needed to support such types of exotic phases \cite{Jiang2012, Yang2012, Moeller2015}.

\begin{figure}
\centering
\includegraphics[width=0.35\textwidth]{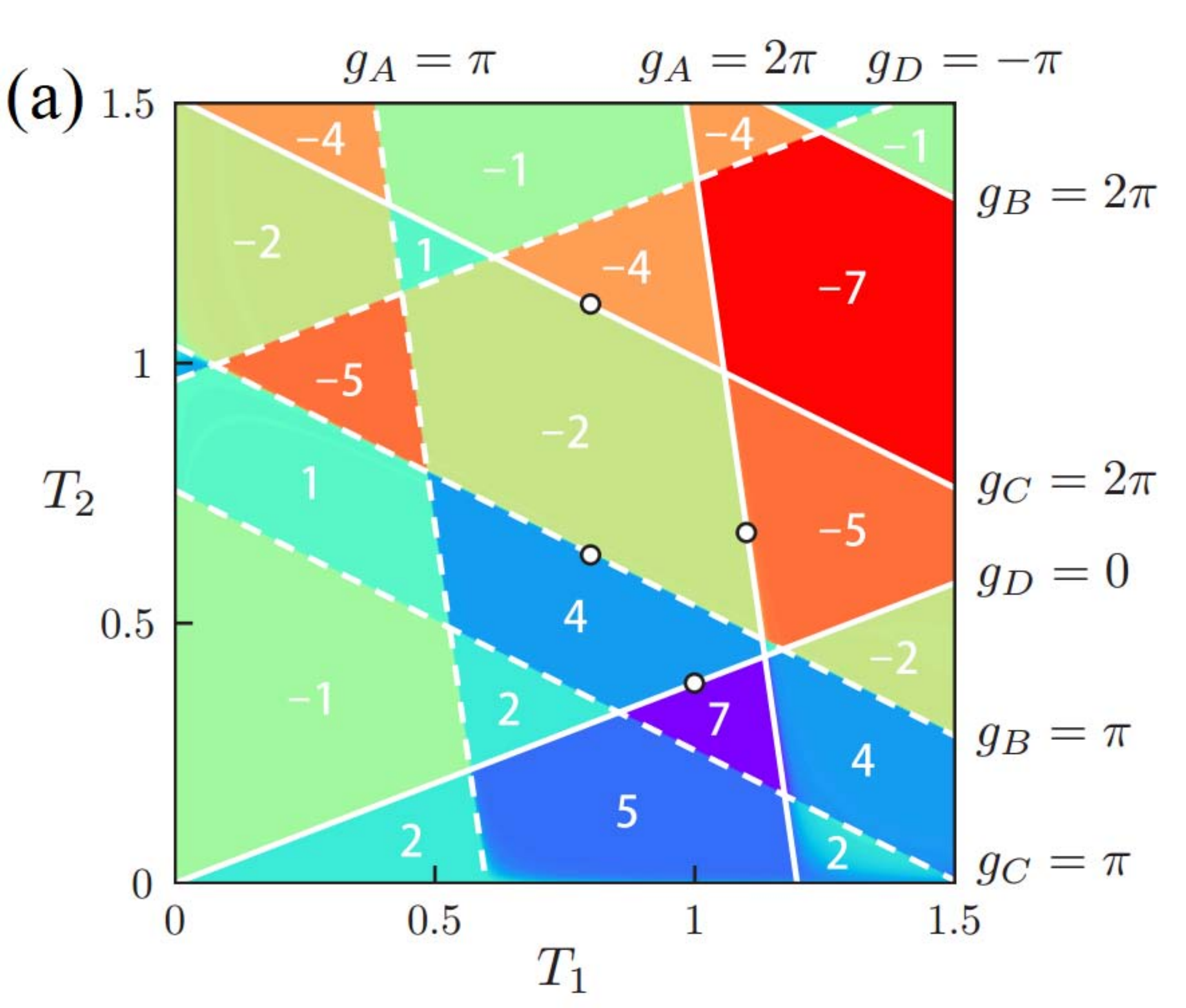}~~\includegraphics[width=.55\textwidth]{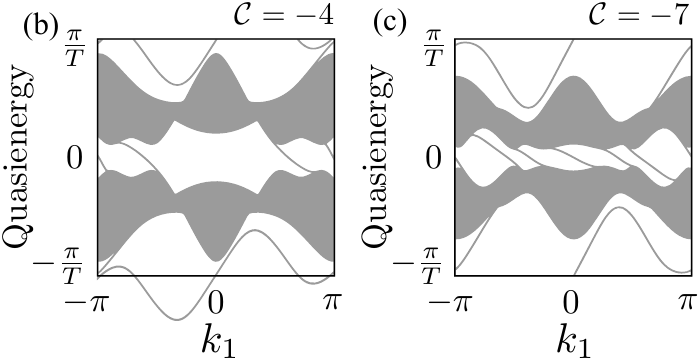}\\
\caption{(a) Phase diagram in the $T_1$-$T_2$ plane obtained by calculating Chern numbers. The phase-boundary lines are analytically obtained from Eqs. \eqref{gen} and \eqref{hh1}. Quasienergy spectra of the periodically Haldane model with $\mathcal{C}=-4$ when $(T_{1},T_{2})=(0.9,1.2)$ in (b) and with $\mathcal{C}=-7$ when $(T_1,T_2)=(1.3,1.2)$] in (c). Parameters $t_2 = 0.8t_1$ and $M = 0$ are used. Reproduced
figures from \cite{Xiong2016}. }
 \label{phase}
\end{figure}

Periodic driving widely used in graphene \cite{McIver2020}, optical lattices \cite{Eckardt2017}, and photonic crystals \cite{Cheng2019} supplies a useful way to create the large-Chern-number topological phases \cite{Xiong2016}. Consider the Haldane model with up to the third-neighbor hopping included. In the operator basis $C_{\mathbf{k}}=(c_{a\mathbf{k}},c_{b\mathbf{k}})^{T}$, its static Hamiltonian can be written as $H=\sum_{\mathbf{k}\in \text{BZ}}C_{\mathbf{k}}^{\dag}[\mathbf{h}(\mathbf{k})\cdot{\pmb\sigma}]C_{\mathbf{k}}$. The associated Bloch vectors $\mathbf{h}(\mathbf{k})$ are
\begin{align}\label{hamiltonian}
\begin{split}
&h_{x}(\mathbf{k})=t_{1}[1+\cos(\mathbf{k}\cdot\mathbf{a}_{1})+\cos(\mathbf{k}\cdot\mathbf{a}_{2})]+t_{3}\{2\cos[\mathbf{k}\cdot(\mathbf{a}_1-\mathbf{a}_2)]\\&~~~~~~~~~~+\cos[\mathbf{k}\cdot(\mathbf{a}_1+\mathbf{a}_2)]\},\\
&h_{y}(\mathbf{k})=t_{1}[\sin(\mathbf{k}\cdot\mathbf{a}_{1})+\sin(\mathbf{k}\cdot\mathbf{a}_{2})]+t_{3}\sin[\mathbf{k}\cdot(\mathbf{a}_1+\mathbf{a}_2)],\\
&h_{z}(\mathbf{k})=2t_{2}\sin\phi\{\sin(\mathbf{k}\cdot\mathbf{a}_1)-\sin(\mathbf{k}\cdot\mathbf{a}_2)-\sin[\mathbf{k}\cdot(\mathbf{a}_1-\mathbf{a}_2)]\}+M,
\end{split}
\end{align}
where $t_{j}$ are the $j$th-neighbor hopping amplitude, $\phi$ is the phase gained by $t_{2}$ hoppings, and $M$ is the mass term. The primitive vectors are $\mathbf{a}_1=(\sqrt{3}/2,3/2)$ and $\mathbf{a}_2=(-\sqrt{3}/2,3/2)$ in the unit of the hexagonal lattice constant \cite{Haldane1988}. Figure \ref{phase} shows the phase diagram in the $T_1$ and $T_2$ plane by periodically switching $t_{3}=0.75t_1$ and $\phi=-\pi/6$ for duration $T_1$ to $t_{3}=-0.75t_1$ and $\phi=-\pi/2$ for duration $T_2$. A widely tunable Chern numbers with the equal numbers of edge states are generated [see Figs. \ref{phase}(b) and \ref{phase}(c)]. Due to the high controllability of the cold atom system \cite{Juenemann2017,Potirniche2017}, a periodic driving scheme was proposed to realize large-topological-number phases via periodically changing the optical lattices \cite{Liu2019}.

\subsubsection{Non-Hermitian Floquet topological phases}
Inspired by the recent experimental realization of topological insulators and the rapid development of non-Hermitian photonics, non-Hermitian topological phases have attracted much attention. From the fundamental physics aspect, people desire to know can the well-developed topological theory in Hermitian systems be generalized to the non-Hermitian cases. This is nontrivial because the skin effect of the non-Hermitian system could cause the breakdown of the bulk-boundary correspondence, which lays the foundation in the topological description in Hermitian systems. From the application aspect, non-Hermitian topological phases have exhibited significant advantages in diverse applications, e.g., laser, invisible media, and sensing \cite{Yuto}. In order to prompt the application of non-Hermitian topological phases, the creation of novel non-Hermitian topological phases in a controllable way via Floquet engineering the system by periodic driving is highly promising. However, the interplay between the non-Hermitian skin effect and periodic driving makes it very hard to characterize the non-Hermitian Floquet topological phases. A recent study has established a complete description of the non-Hermitian Floquet topological phases \cite{Wu2020}.

Consider the non-Hermitian Su-Schrieffer-Heeger model
\begin{eqnarray}
H=\sum_{l=1}^L[(t_1+\frac{\gamma}{2})a^{\dag}_lb_l+(t_1-\frac{\gamma}{2})b^{\dag}_la_l+t_2(a^{\dag}_lb_{l-1}+\text{h.c.})],~~~\label{Hmat}
\end{eqnarray}where $a_l$ ($b_l$) are the annihilation operators on the sublattice $A$ ($B$) of the $l$th lattice, and $L$ is lattice length. In momentum space and the operator basis $(\tilde{a}_k,\tilde{b}_k)^T$ with $\tilde{a}_k$ ($\tilde{b}_k$) being the Fourier transform of $a_l$ ($b_l$), it reads $\mathcal{H}(k)=d_x\sigma_x+(d_y+i{\gamma}/{2})\sigma_y$, where $d_x=t_1+t_2\cos k$ and $d_y=t_2\sin k$. The bands close at $k=\pi$ (or $0$) when $t_1=t_2\pm\gamma/2$ (or $-t_2\pm\gamma/2$). It is in conflict with the result under the open-boundary condition, where the bands close when $t_1=\sqrt{t_2^2+\gamma^2/4}$. It is called the non-Hermiticity caused breakdown of bulk-boundary correspondence \cite{Lee2016, Yao2018}. The problem for the static system with chiral symmetry $\sigma^{-1}_z\mathcal{H}(k)\sigma_z=-\mathcal{H}(k)$ is cured via replacing $e^{ik}$ by $\beta=\sqrt{\lvert \frac{t_1-\gamma/2}{t_1+\gamma/2}\lvert}e^{ik}$. Then the Hamiltonian is converted into $\mathcal{H}(\beta)=\sum_{n=\pm}R_{n}(\beta)\sigma_{n}$ with $\sigma_{\pm}=(\sigma_x\pm i\sigma_y)/2$ and $R_{\pm}(\beta)=t_1\pm\frac{\gamma}{2}+\beta^{\mp1} t_2$. Here $\beta$ defines a generalized Brillouin zone (GBZ). Its topological property is described by the winding number
$\mathcal{W}=-(\mathcal{W}_{+}-\mathcal{W}_{-})/2$, where $\mathcal{W}_{\pm}=\frac{1}{2\pi}[\arg R_{\pm}(\beta)]_{C_{\beta}}$ with $[\arg R_{\pm}(\beta)]_{C_{\beta}}$ are the phase change of $R_{\pm}$ as $\beta$ counterclockwisely goes along the GBZ $C_{\beta}$ \cite{Yokomizo2019}. When $|t_1|<\sqrt{t^2_2+\gamma^2/4}$, $\mathcal{W}=1$ and a pair of edge states is formed.

\begin{figure}
\centering
\includegraphics[height=0.35\textwidth]{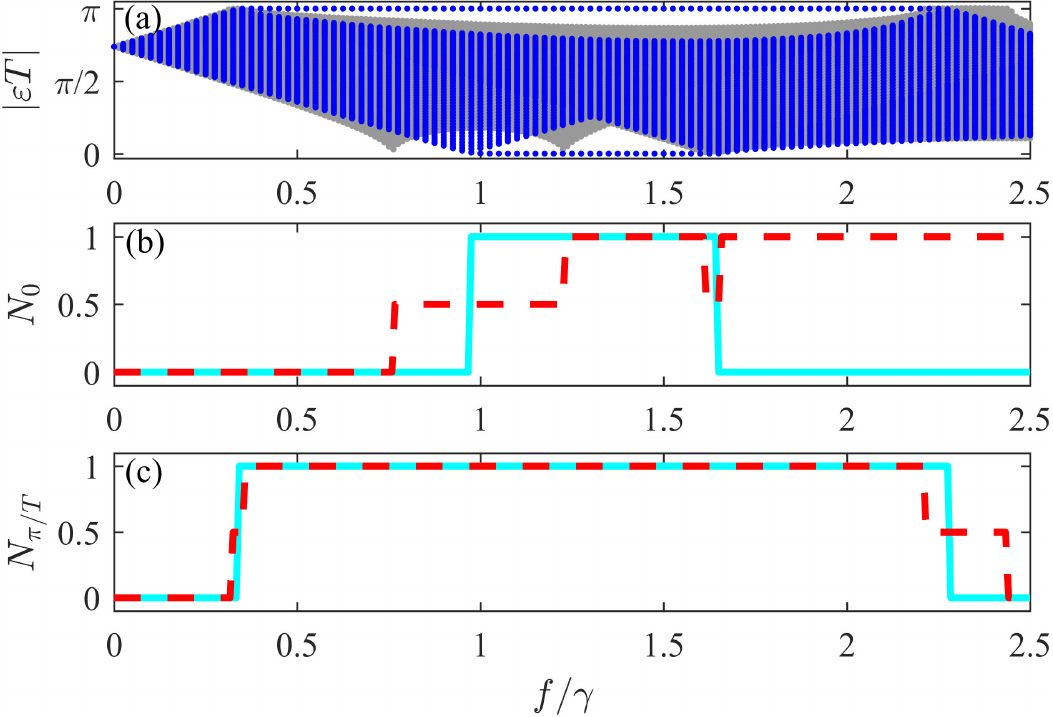}~~\includegraphics[height=.35\textwidth]{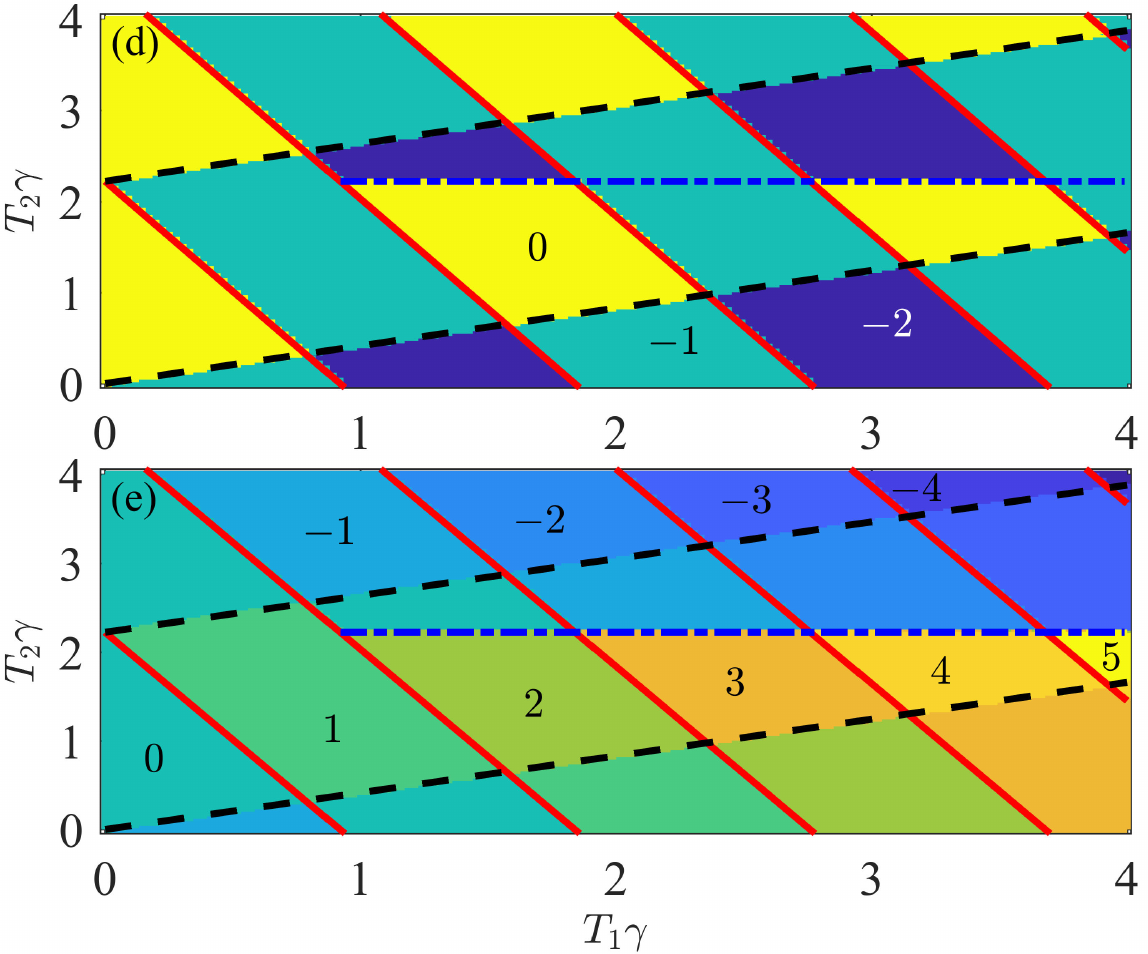}\\
\caption{(a) Quasienergy spectra with the change of the driving amplitude under the open (blue lines) and periodic (gray lines) boundary conditions. Numbers of $0$-mode (b) and $\pi/T$-mode (c) edge states defined in the conventional (red dashed) and generalized (cyan solid) BZ. Parameters $t_1=2.0\gamma$, $T_1=T_2=0.6\gamma^{-1}$, $q=3.0$, and $L=80$ are used in (a), (b), and (c). Phase diagram characterized by $\mathcal{W}_1$ (d) and $\mathcal{W}_2$ (e). The phase boundary lines are obtained from Eqs. \eqref{gen} and \eqref{hh1}. Parameters $t_1=1.5\gamma$, $f=2\gamma$, and $q=0$ are used in (d) and (e). Reproduced
figures from \cite{Wu2020}.}
 \label{nhmf}
\end{figure}

If the periodic driving
\begin{equation}
t_2(t) =\begin{cases}f ,& t\in\lbrack mT, mT+T_1)\\q\,f,& t\in\lbrack mT+T_1, (m+1)T), \end{cases}~m\in \mathbb{Z}, \label{procotol}
\end{equation}is applied to the system, then the chiral symmetry is broken, which makes the GBZ method well developed for the chiral symmetric system not directly applicable. To resolve this problem, two similarity transformations $G_{j}=e^{i(-1)^j\mathcal{H}_jT_j/2}$ are introduced to covert the one-period evolution operator $U_T$ into $\tilde{U}_{T,1}=e^{-i\mathcal{H}_1T_1/2}e^{-i\mathcal{H}_2T_2}e^{-i\mathcal{H}_1T_1/2}$ and $\tilde{U}_{T,2}=e^{-i\mathcal{H}_2T_2/2}e^{-i\mathcal{H}_1T_1}e^{-i\mathcal{H}_2T_2/2}$, from which two effective Hamiltonian $\tilde{H}_{\text{eff},j}={i\over T}\ln \tilde{U}_{T,j}$ can be defined. It can be analytically proved that the chiral symmetry is recovered in $\tilde{H}_{\text{eff},j}$. Then after introducing the GBZ $\beta$, two winding numbers $\mathcal{W}_j$ associated with $\tilde{\mathcal{H}}_{\text{eff},j}$ can be calculated. The numbers of zero- and $\pi/T$-mode edge sates relate to $\mathcal{W}_j$ as $N_0=|\mathcal{W}_1+\mathcal{W}_2|/2$ and $N_{\pi/T}=|\mathcal{W}_1-\mathcal{W}_2|/2$ \cite{Asboth2014}. Figure \ref{nhmf}(a) shows the quasienergy spectrum under the open-boundary conditions. It is different from the quasienergy spectrum under the periodic-boundary condition, which is called the breakdown of bulk-boundary condition. Although qualitatively capturing the exceptional points of the quasienergy under the periodic-boundary condition, the ill-defined topological numbers from the conventional BZ nonphysically take half integers [see the dashed lines in Fig. \ref{nhmf}(b) and \ref{nhmf}(c)]. However, the ones from the GBZ correctly count the number of the edge states obtained in the open-boundary condition [see the solid lines in Fig. \ref{nhmf}(b) and \ref{nhmf}(c)]. Figures \ref{nhmf}(d) and \ref{nhmf}(e) show the phase diagram in the $T_1$-$T_2$ plane. A widely tunable number of non-Hermitian edge states are induced by changing the driving parameters.

All the above results reveal that, without changing the intrinsic parameters in the static system, the periodic driving supplies us another control dimension to adjust the topological edge states. This is useful in
the application of topological physics.

\section{Conclusion and outlook} \label{sec:conclu}

In summary, we have reviewed the recent progress on quantum control in open and periodically driven systems. Particular attention has been paid to the distinguished role of bound states in suppressing decoherence of open systems and realizing exotic topological phases in periodically driven systems. It has been revealed that the decoherence of the open system connects tightly with the feature of the energy spectrum of the total system consisting of the open system and its environment. With the formation of a bound state, the decoherence of the open system would be suppressed. It has a profound impact on entanglement preservation, quantum speedup, noncanonical thermalization, quantum metrology, and quantum plasmonics. A parallel result can be established in a periodically driven open system, where the formation of FBSs in the quasienergy spectrum can suppress the dissipation of the open system and the aging of the dissipative quantum battery. It is interesting to see that the controllability of periodic driving on engineering the FBSs can be applied to realize diverse exotic topological phases, e.g. the widely tunable numbers of Majorana modes, Chern insulators, and non-Hermitian topological insulators. The merit of such types of topological phases induced by periodic driving resides in that their controllability is dramatically enhanced because time as a novel control dimension is introduced to the system by periodic driving. It might inspire the exploration of their application via tuning the number of edge/surface states by the periodic driving.
The bound state and its distinguished role in the static open-system dynamics have been observed in circuit QED \cite{Hood2016,Liu2017} and ultracold atom systems \cite{Kri2018}. These progresses provide strong support to the bound-state mechanism and indicate that the profound consequences of the bound reviewed in this article is realizable in the current state of the art. Such experimental feasibility paves the way to beat the decoherence in diverse protocols in quantum technology.

The studies presented in this article hopefully open a number of research lines that deserve to be explored in the future. First, the combination of quantum optics and topological physics has emerged a field called topological quantum optics \cite{PhysRevLett.124.083603,Barik666,Belloeaaw0297,PhysRevLett.125.163602}, where the environments interacting with the quantum emitters have topologically nontrivial energy bands. It has potential application in a quantum network. Interesting questions include: Can the system-environment bound states be formed and how the interplay of these bound states with the intrinsic edge states of the environment affects the dynamics of the quantum emitters? Second, parallel to the bound states in the bandgap, the bound states in the continuum are also widely studies \cite{PhysRevLett.107.183901,PhysRevLett.122.073601}. A recent study has showed that the bound states in the continuum can be used to realize the higher-order topological phases \cite{PhysRevLett.125.213901}. Can they be efficietly manipulated by periodic driving such that more colorful higher-order topological phases are generated in a controllable manner? Third, time crystal is a novel state of matter showing the spontaneous symmetry breaking in the time domain of a quantum system under periodic driving \cite{PhysRevLett.118.030401,Choi2017}. It arises from the competition of many-body localization due to the disorder against Floquet thermalization under periodic driving \cite{PhysRevB.93.104203,PhysRevB.97.245122}. Inspired by the dominated role of bound states in noncanonical thermalization in static systems, a natural question is: Does the avoidance of Floquet thermalization in order to form time crystal is due to the presence of FBSs in quasienergy spectrum? The answer to this question might give the intrinsic physical mechanism to the presence of time crystal from the quasienergy-spectrum feature.

\section*{Acknowledgments}
The work is supported by the National Natural Science Foundation of China (Grant Nos. 11875150 and 11834005).

\bibliographystyle{unsrt}
\bibliography{BSF}

\end{document}